\documentclass[a4paper]{spie}  

\usepackage[]{graphicx}

\def\vereq#1#2{\lower3.75pt\vbox{\baselineskip0.5pt \lineskip0.5pt
\ialign{$#1\hfill##\hfil$\crcr#2\crcr\sim\crcr}}}

\def\gtrsim{\mathrel{\mathpalette\vereq>}}

\def\agt{\gtrsim}

\title{Electronic reconstruction in correlated electron heterostructures} 

\author{Satoshi Okamoto and Andrew J. Millis 
\skiplinehalf
Department of Physics, Columbia University, 538 West 120th Street, New York, NY 10027
}

\authorinfo{Further author information: 
(Send correspondence to S.O.) E-mail: okapon@phys.columbia.edu, Telephone: +1-212-854-6712}
 
\begin{document} 
  \maketitle 

\begin{abstract}
Electronic phase behavior in correlated-electron systems is a fundamental problem of 
condensed matter physics. 
We argue here that the change in the phase behavior near the surface and interface, 
i.e., {\em electronic reconstruction}, 
is the fundamental issue of the correlated-electron surface or interface science. 
Beyond its importance to basic science, understanding of this behavior is crucial 
for potential devices exploiting the novel properties of the correlated systems. 
We present a general overview of the field, and then illustrate the general concepts 
by theoretical studies of the model heterostructures comprised of 
a Mott-insulator and a band-insulator, which show that 
spin (and orbital) orderings in thin heterostructures are generically different 
from the bulk and that the interface region, about three-unit-cell wide, is always metallic, 
demonstrating that {\em electronic reconstruction} generally occurs. 
Predictions for photoemission experiments are made 
to show how the electronic properties change as a function of position, 
and the magnetic phase diagram is determined as a function of temperature, 
number of layers, and interaction strength. 
Future directions for research are also discussed. 

\end{abstract}



\section{INTRODUCTION}
\label{sec:intro}

{\em Correlated electron systems} such as transition metal oxides are 
materials in which strong electron-electron or electron-lattice interactions
produces behavior incompatible with the standard {\em density functional plus
Migdal-Eliashberg theory} which describes most compounds. The past decade
has seen tremendous progress in the physics and materials science of
correlated-electron systems. Significant improvements in crystal and film
growth, in measurement techniques and in theory have led to a much improved
understanding of the bulk properties of these materials. An important
finding is that correlated electron systems exhibit a multiplicity of
interesting phases (superconducting, magnetic, charge, and orbitally ordered)
along with associated excitations. For recent reviews, see Ref.~%
\citenum{Imada98}, or the articles in Ref.~\citenum{Tokura00}.

The recent success in treating bulk properties suggests that the time is
ripe for a systematic study of the surface and interface properties of
correlated electron systems. In addition to its basic importance as a
fundamental question in materials science, correlated electron
surface/interface science should provide the necessary scientific background
for study of potential devices exploiting correlated electron properties,
because essentially any device must be coupled to the rest of the world via 
motion of electrons through an interface. 
The fundamental interest of bulk correlated electron materials lies in novel phases they exhibit,
and we therefore suggest that the fundamental issue for the nascent field of 
{\em correlated electron surface science} is {\em how does the electronic phase
at the surface or interface differ from that in the bulk;} in other words, 
{\em what is the electronic surface reconstruction.}

As in the ordinary surface or interface science, 
many physics and material science issues arise in considering the behavior
of correlated electrons near surfaces and interfaces. 
{\em Atomic reconstruction} may occur, and may change the underlying electronic structure. 
For example, the authors of Ref.~\citenum{Moore04} argue that, in two-dimensional ruthenates, 
a change in tilt angle of the surface RuO$_6$ octahedra increases the electronic hopping, 
thereby allowing the metallic phase to persist to lower $T$ than in the bulk. 
{\em Reduced coordination number} at surfaces is supposed to enhance correlation effects 
as discussed by Potthoff and Nolting~\cite{Potthoff99a,Potthoff99b}, 
Schwieger {\it et al.}~\cite{Schwieger03}, and Liebsch~\cite{Liebsch03}. 
Also, Hesper and co-workers have shown that the [111] surface of K$_3$C$_{60}$ differs from the bulk. 
They noted that a change in structure will lead to {\em changes in Madelung potentials}, 
and to the screening which helps define the values 
of many-body interaction parameters.\cite{Hesper00} 
{\em Leakage of charge} across an interface may change densities away from the commensurate values 
required for insulating behavior. 

On the experimental side, 
a variety of heterostructures have been fabricated and studied. 
For the surface, i.e., interface between a material and a vacuum,  
Maiti and collaborators have shown that the surface and bulk electronic phases of 
Ca$_{1-x}$Sr$_x$VO$_3$ are significantly different.\cite{Maiti01} 
They argued that the enhanced incoherent part of the photoemission spectra 
may be due to the reduced coordination number at the surface and/or surface reconstruction. 
Moore and collaborators have demonstrated that in the correlated electron
system Ca$_{1.9}$Sr$_{0.1}$RuO$_4$ (which exhibits a Mott metal-insulator 
transition), the surface layers remain metallic down to a lower temperature
than does the bulk system.\cite{Moore04} 
We have proposed that in this system 
electron-lattice coupling is crucial for the Mott metal-insulator transition.\cite{Okamoto04b} 
Therefore, the surface Mott transition in these materials may be different from that in the bulk; 
at the surface, due to the larger spring constant (estimated from the higher phonon frequency), 
electron-lattice coupling is reduced, and surface metal transition is suppressed. 

A variety of interfaces between different materials includes 
high-$T_c$ cuprates~\cite{Ahn99,Gariglio02,Schneider02}, 
Mott-insulator and band-insulator heterostructure~\cite{Ohtomo02}, 
and superlattices of transition-metal oxides.%
\cite{Izumi99,Izumi01,Biswas00,Biswas01,Warusawithana03,Bowen03,Nakagawa05}
Izumi and co-workers have fabricated {\em digital heterostructures} composed of different transition
metal oxides and have demonstrated changes in electronic phase and other
properties depending on the thicknesses of different layers.\cite{Izumi99}
Warusawithana and collaborators fabricated and measured 
a variety of superlattices of dielectric materials with broken inversion symmetry.\cite{Warusawithana03} 
In an experimental tour-de-force, Ohtomo and co-workers have
demonstrated the fabrication of atomically precise digital heterostructures
involving a controllable number $n$ of planes of LaTiO$_3$ (a
correlated-electron Mott-insulating material) separated by a controllable
number $m$ of planes of SrTiO$_3$ (a more conventional band-insulating
material) and have measured both the variation of electron density
transverse to the planes and the dc transport properties of the
heterostructure.\cite{Ohtomo02} Their work opens the door to controlled
studies both of correlated electron physics in confined dimensions and of
the behavior of the interface between a correlated system and a uncorrelated one. 
Bowen and collaborators fabricated mangnites based tunneling magnetoresistance (TMR) junctions, 
and succeeded in obtaining the high spin polarization as in the bulk materials 
at the lowest temperature.\cite{Bowen03} 
However, magnetoresistance is lost well below the bulk Curie temperature. 
Similarly, Nakagawa and collaborators fabricated manganite-titanate heterojunctions and 
measured their current-bias voltage characteristics and junction capacitance 
with and without the applied magnetic field.\cite{Nakagawa05} 
They observed a significant magnetic-field dependence of the junction capacitance at low temperature 
indicating the change in the electronic state (possibly spin and orbital states) 
of manganites region near the interface under the magnetic field. 
Finally, with an elaborate sample preparation, 
Schneider {\it et al}. were successful in extrapolating the resistance of single grain boundary 
of YBa$_2$Cu$_3$O$_{7-\delta}$.\cite{Schneider02} 
This is clearly supported by the fact that 
the resistance is not affected by the onset of the superconducting transition. 
The resistance is found to decrease linearly with the increase of temperature, and 
it has been suggested that randomly distributed magnetic impurities are responsible 
for the scattering mechanism.\cite{Mannhart05} 

On the theoretical side, there have been several developments. 
The enhanced correlations near the surface due to the reduced coordination 
number\cite{Potthoff99a,Potthoff99b,Schwieger03,Liebsch03} 
could presumably induce surface magnetic ordering, 
this had been discussed in a mean field treatment of 
the Hubbard model by Potthoff and Nolting~\cite{Potthoff95}. 
Matzdorf {\it et al.} proposed that ferromagnetic ordering is stabilized at the surface of 
two-dimensional ruthenates by a lattice distortion~\cite{Matzdorf00}, but this has not yet been observed. 
Calderon {\it et al}. discussed possible surface antiferromagnetism in manganites 
arising from a competition between reduced kinetic energy and antiferromagnetic superexchange interaction
between the nearest-neighboring local spins.\cite{Calderon99}
The effect of bulk strain on the magnetic ordering in perovskite manganites 
was discussed by Fang {\it et al.}~\cite{Fang00} 
Effect of strain had been intensively studied on the ferroelectric materials by using 
the first-principle calculation. (For example, see Refs.~\ \citenum{Tinte03} and \citenum{Bungaro04}.) 
Further, Ederer and Spaldin studied the effects of strain and oxygen vacancy on {\em multiferroicity} 
in bismuth ferrite.\cite{Ederer05} 
Popovic and Satpathy applied LSDA and LSDA+$U$ methods to compute the magnetic properties of 
LaTiO$_3$/SrTiO$_3$ heterostructures fabricated by Ohtomo {\it et al}.\cite{Popovic05} 
Going beyond the study of static properties, 
Freericks applied the dynamical-mean-field method to the correlated [001] heterostructures 
comprised of non-correlated and strongly-correlated regions, 
and computed the conductance perpendicular to the plane.\cite{Freericks04} 
In his model, the correlated region is described by the Falicov-Kimball model, 
which is a simplified version of the Hubbard model 
neglecting the electron hopping integral of one of two spin components, 
and the computations are limited to the particle-hole symmetric case (uniform charge density)
for simplicity. 
Extension to the general model, and in particular, 
to the situation where the charge density is spatially modulated are important future directions. 

Sorting out the different contributions and assessing their effect on the 
many-body physics is a formidable task, which will require a sustained 
experimental and theoretical effort. 
The experiment of Ohtomo \textit{et al.} offers an attractive starting point. 
In this system, the near lattice match (1.5~\% difference in lattice parameter) 
and chemical similarity of the two components (LaTiO$_3$ and SrTiO$_3$) 
suggests that atomic reconstructions, strain, and renormalizations of many-body parameters 
are of lesser importance, so the physical effects of electronic reconstruction can be 
isolated and studied. 
Furthermore, the near Fermi surface states are derived mainly from the Ti $d$ orbitals,\cite{Saitoh95} 
and correspond to narrow bands well described by tight-binding theory. 
Therefore, the model calculation of heterostructures of the type created by Ohtomo {\it et al}. 
is expected to be a good starting point toward a general understanding of 
the correlated electron surface and interface problem. 

In the rest of this paper we review our work on theoretical analysis of the correlated electron
behavior to be expected in lattice-matched digital heterostructures of the
type created by Ohtomo \textit{et al}.\cite{Ohtomo02} 
We focus on electrons in the Ti-derived $d$-bands and include the effects of the long-ranged electric
fields arising both from the La atoms and the electronic charge distribution. 
Ti $d$-bands are represented by either a realistic $t_{2g}$ three-band Hubbard model or 
a simplified single-band Hubbard model. 
The Hartree-Fock (HF) approximation is applied to treat the on-site interactions of the three-band model. 
We calculate the electronic phase diagram as a function of 
on-site interaction parameter and number of La layers and for the relevant 
phases determine the spatial variation of charge, spin and orbital densities.\cite{Okamoto04Nat,Okamoto04PRB1} 
We obtain a complex set of behaviors depending on interaction strength and number of La layers. 
Generally, we find a crossover length of approximately three unit cells, 
so that units of six or more LaTiO$_3$ layers have a central region 
which exhibits bulk-like behavior. 
The outermost $\sim3$ layers on each side are however metallic 
(in contrast to the insulating behavior of bulk LaTiO$_3$ and SrTiO$_3$). 
For very thin superlattices the ordering patterns differ from that in bulk. 
While the HF approximation provides the correct tendency towards 
magnetically and orbitally ordered states, 
this method is known to be an inadequate representation of strongly correlated materials, 
and in particular, does not include the physics associated with proximity to the Mott insulating state. 
Therefore, as a step to go beyond HF approximation, 
we apply the dynamical-mean-field method,\cite{Georges96} which 
provides a much better representation of the electronic dynamics associated 
with strong correlations, to a simplified single-band model, and 
investigate the dynamical properties of correlated-electrons.\cite{Okamoto04PRB2,Okamoto05} 
The dynamical-mean-field analysis confirms the important results obtained by the HF analysis
, i.e., different phases in thin heterostructures than in the bulk and metallic edge, 
but provides significant improvement over the HF results 
for the emergence of the metallic behavior, the magnetic transition temperatures and 
order parameter distributions. 

The rest of this paper is organized as follows: 
In Sec.~\ref{sec:model}, we present our theoretical models and methods applied, 
Sec.~\ref{sec:HF} presents results of the HF analysis of the realistic three-band model,  
and Sec.~\ref{sec:DMFT} presents results of the dynamical-mean-field analysis of the single-band model. 
Section~\ref{sec:conclusion} is devoted to conclusion and discussion. 
Part of the results presented in this paper has already been published in 
Refs.~\citenum{Okamoto04Nat,Okamoto04PRB1}, and \citenum{Okamoto04PRB2}, 
and can be seen in Ref.~\citenum{Okamoto05}.  

\section{Model and method} 
\label{sec:model}

\begin{figure}
\begin{minipage}{7cm}
\begin{center}
\begin{tabular}{c}
\includegraphics[height=4.5cm,clip]{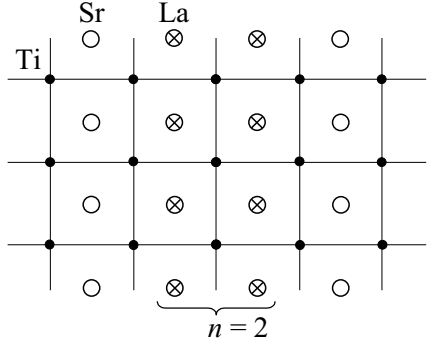}
\end{tabular}
\end{center}
\end{minipage}
\begin{minipage}{9cm}
\caption
{ \label{fig:model} 
Schematic figure of the model used in the present study. Open and
crossed circles show the positions of Sr and La ions. (La layer number $n=2$%
) The dots show the positions of Ti ions. $x,y$ axes are chosen to be
parallel to the La plane, and $z$ axis is perpendicular to the plane.}
\end{minipage}
\end{figure}

In this study, we consider an infinite crystal of SrTiO$_3$, into which $n$ adjacent [001] planes 
of LaTiO$_3$ have been inserted perpendicular to one of the Ti-Ti bond directions, 
as shown in Fig.~\ref{fig:model} ([001] heterostructure). 
We choose the $z$ direction to be perpendicular to the LaTiO$_3$ planes, 
so the system has (discrete) translation symmetry in the $xy$ direction. 
Both LaTiO$_3$ and SrTiO$_3$ crystallize in the simple $AB$O$_3$ perovskite
structure \cite{Maclean79,Sunstrom92} (more precisely, very small distortions occur) 
and as noted by Ref.~\citenum{Ohtomo02} the lattice constants 
of the two materials are almost identical; $a_{LaTiO_3} 
\simeq a_{SrTiO_3} = 3.9$~\AA, minimizing structural discontinuities at the 
interface and presumably aiding in the growth of high quality digital heterostructures.
Therefore, in this study, we neglect lattice distortions 
and focus on the purely electronic model. 
Possible effects of lattice distortions will be briefly discussed below. 

We consider the following two model heterostructures: 
(1) realistic three-band model and (2) single-band model. 
We apply the HF approximation to 
the three-band model and discuss static properties such as spin and orbital orderings 
of the heterostructures. 
The single-band heterostructure is not a fully realistic representation of the systems studied 
Ref.~\citenum{Ohtomo02}, but the essential physics associated with 
{\em charge leakage} and {\em strong correlation} is included. 
The single-band model is analyzed using the dynamical-mean-field theory (DMFT). 

Applying beyond-HF methods including the DMFT to the multi-band model is highly desirable, but 
this is beyond the current computational ability. 
In order to apply the DMFT to the heterostructure problem, 
it is required to solve many quantum impurity models coupled with each other self-consistently. 
Further, numerically expensive method is usually applied to solve the impurity model 
such as the quantum Monte-Carlo (QMC) method or the exact-diagonalization (ED) method. 
However, as will be discussed later, 
interesting physics appears at strong coupling regime, 
the interaction strength is larger than the band width, and at low-but-non-zero temperature, 
lower than the magnetic ordering temperature of the bulk material. 
This regime is not easily accessible either by QMC and by ED. 
In order to apply the DMFT method to realistic multi-band models, 
numerically inexpensive but reliable impurity solvers are required. 
Work in this direction is still in progress. 

In both the models, the hamiltonian $H_{tot}$ is comprised of three terms: 
$H_{tot}=H_{hop}+H_{on-site}+H_{Coul}$ with 
electron hopping term $H_{hop}$, on-site interaction term $H_{on-site}$, 
and long-ranged Coulomb interaction term $H_{Coul}$. 
The hopping term $H_{hop}$ and the on-site interaction term $H_{on-site}$ 
vary depending on the model one considers as presented in the following subsections, 
while the long-ranged Coulomb interaction term $H_{Coul}$ has the same form 
and consists of two terms: (i) the Coulomb force arising from the extra charge 
on the La relative to Sr defining the heterostructure, 
and (ii) the Coulomb force arising from the electrons on other Ti sites. 
Thus, 
\begin{eqnarray}
H_{Coul} = \sum_i V (\vec r_i) n_{tot} (\vec r_i) \label{eq:Hcoul}
\label{eq:Hcoul}
\end{eqnarray}
with 
\begin{eqnarray}
V (\vec r_i) = - \!\! \sum_{\mbox{\scriptsize La-sites} \atop {j}} 
\frac{e^2}{\varepsilon |\vec r_j^{La} - \vec r_i|}
+ \frac{1}{2} \sum_{\mbox{\scriptsize Ti-sites} \atop {j \ne i}}
\frac{e^2 n_{tot} (\vec r_j)}{\varepsilon |\vec r_j - \vec r_i|}
\label{eq:Vc}
\end{eqnarray}
and 
$n_{tot} (\vec r_i) = \sum_{\sigma (\alpha)} d_{i (\alpha) \sigma}^\dag d_{i (\alpha) \sigma}$ 
(sum over the orbital index $\alpha$ is necessary for the three-band model). 
Here the $\vec r_i$ are the positions of the Ti$^{(B)}$ sites in the $AB$O$_3$
lattice and $\vec r_j^{La}$ label the actual positions of the La ions, which
reside on the $A$ sites. 

We denote the dielectric function of the host lattice by $\varepsilon$. An
interesting issue arises here: SrTiO$_3$ is a nearly ferroelectric material.%
\cite{Sakudo71} The static dielectric constant becomes very large at long
wavelength and low temperatures, but $\varepsilon$ is much smaller at high
frequencies, room temperature, or short length scales. Also the polarization 
$P$ will increase more slowly at higher fields, and relevant quantity is $P/E$. 
In our work, we have chosen $\varepsilon \sim 15$ as a compromise between these effects, 
and have chosen a dimensionless parameter for the long-ranged Coulomb interaction 
$E_c = e^2/(\varepsilon a t)=0.8$ corresponding to 
a lattice constant $a=3.9$~\AA \, transfer intensity $t=0.3$~eV.  
The effect of different choices of $\varepsilon$ will be discussed later. 

In addition to the strong ferroelectric tendency, 
other issues have been neglected in theoretical work so far. 
We emphasize that incorporating these effects 
in a more realistic manner is a important question for future research. 
Important effects include 
(I) Change in the electron transfer intensity: 
As can be seen in the TEM image of Fig.~1 in Ref.~\citenum{Ohtomo02}, 
Ti sites do not show significant displacement throughout the heterostructure. 
Thus, change in the transfer intensity between neighboring Ti $d$-orbitals 
$\delta t$ is given by $\delta t \propto \delta^2$ 
with $\delta$ the oxygen displacement. Therefore, it is expected be small. 
However, when a more realistic model including O $p$-orbitals is considered, 
this effect can not be neglected because the Ti--O transfer is changed linearly in $\delta$. 
Further, 
(II) Change in the on-site interaction parameters: 
Due to the lattice distortion, screening effect might be different, thus, changing the interaction parameters. 
(III) Degeneracy lifting in the Ti $d$-level: 
Due to the small but non-zero mismatch in the lattice constants, 
$a=3.91$~\AA \ for SrTiO$_3$ substrate, and $a=3.97$~\AA \ for LaTiO$_3$,\cite{Sunstrom92} 
The Ti $d$-orbitals in LaTiO$_3$ region would suffer from an in-plane compressive strain, and 
the energy levels of the orbitals elongated in the plane would be raised. 
(IV) Absence of inversion symmetry: 
This would admix the Ti $3d$ $t_{2g}$-orbital with the orbitals with the different symmetry including 
the Ti $3d$ $e_g$- and $4p$-orbitals via Ti--O hybridization. 
(V) Chemical effect: 
Because of the chemical difference between the two compounds, there may be 
additional level difference in conduction band on top of the one from the charge difference 
between La and Sr ions, which is considered here. 

Before presenting the explicit form of $H_{hop}$ and $H_{on-site}$, 
let us discuss a difficulty one encounters when analyzing the hamiltonian $H_{tot}$, 
in which the detail of $H_{hop}$ and $H_{on-site}$ does not matter. 
As can be seen from the geometry of the heterostructures we concern, 
our theoretical model is essentially equivalent to a two-dimensional quantum well. 
Therefore, there are two types of solutions to the one-electron equations: 
bound states, which decay as $|z| \rightarrow \infty$, and continuum states, which do not. 
The bound states give rise to sub-bands, some of which are partially occupied. 
The ground state is obtained by filling the lowest sub-bands up to 
the appropriate chemical potential (determined by charge neutrality); 
the interaction-related terms such as the electron self-energy are then recomputed 
and the procedure is repeated until self-consistency is obtained. 
Charge neutrality requires that the total density of electrons in 
the bound-state sub-bands equals the total density of La ions. 
However, the interplay between electron-La attraction and electron-electron repulsion 
leads (in almost all of the cases we have studied) to a very weak binding of 
the highest-lying electron states; indeed for large $U$ the Fermi level of 
the partially filled sub-bands is only infinitesimally below the bottom of the continuum bands.
Therefore, a large number of iterations is required to obtain accurate self consistency. 
In the HF analysis for the three-band model (basically $T=0$), 
it is required to iterate 100--1000 times 
(the largest number of iterations is necessary for the phases with complicated spin and orbital orderings). 
In the DMFT analysis for the single-band model, 50--500 iterations are required. 
Again, the largest number of iterations is required for the magnetic phase at low temperature; 
fewer (20--50) are needed for the non-magnetic one at high temperature. 
Supercell techniques may alleviate this problem, at the cost a less precise treatment of the charge tails.

In the following subsections, we present $H_{hop}$ and $H_{on-site}$ for the three-band and
the single-band models. We also present the numerical methods employed to analyze these models. 

\subsection{Three-band model}

The relevant electronic orbitals are derived from the Ti $t_{2g}$-symmetry $d$-states, 
and may be labeled as $d_{xy}, d_{xz}, d_{yz}$. 
The derived bands\cite{Fujitani95} for bulk materials are to a good approximation 
given by a nearest-neighbor tight binding model with hopping parameter of magnitude 
$t \simeq 0.3$~eV and spatial structure given by the Slater-Koster formula,\cite{Slater54} 
so the $d_{xy}$ states disperse only in the $xy$ plane \textit{etc}. 
We take the form of the on-site interactions determined by Mizokawa \textit{%
et al}.\cite{Mizokawa95} and adopt values as discussed below. 
Thus, $H_{hop} = \sum_\alpha H_{hop}^{(\alpha)}$ with 
\begin{eqnarray}
H_{hop}^{(xy)}= -2t \sum_{\vec k \sigma} (\cos k_x + \cos k_y) 
d_{xy \, \sigma \vec k}^\dag d_{xy \, \sigma \vec k} \label{eq:Hhop}
\end{eqnarray}
and similarly for $xz, yz$. 
The onsite $H_{on-site}$ is 
\begin{eqnarray}
H_{on-site}=\sum_i \biggl\{ U \sum_\alpha n_{i \alpha \uparrow} n_{i \alpha \downarrow}
+(U^{\prime}-J) \sum_{\alpha > \beta, \sigma} n_{i \alpha \sigma} n_{i \beta \sigma}  
+U^{\prime}\sum_{\alpha \ne \beta} n_{i \alpha \uparrow} n_{i \beta \downarrow} 
+ J \sum_{\alpha \ne \beta} d_{i \alpha \uparrow}^\dag d_{i \beta \uparrow} 
d_{i \beta \downarrow}^\dag d_{i \alpha \downarrow} \biggr\}.
\label{eq:onsite3}
\end{eqnarray}
We have omitted a pair-transfer ($J'$) interaction which does not affect our results. 
For definiteness we follow other studies which employ the ratios 
$U^{\prime}=7U/9$ and $J=U/9$ which are very close to those determined by 
Mizokawa.\cite{Mizokawa95} 
Many workers have used the value $U \sim 5\mbox{--}6$~eV~$\sim 18t \mbox{--}20t$ 
estimated from high energy spectroscopies.\cite{Saitoh95} 
However, optical conductivity studies of LaTiO$_3$ and related compounds
such as YTiO$_3$ find rather small gaps, in the range 0.2--1~eV,\cite%
{Okimoto95} suggesting $U \sim 2.5$~eV~$\sim 8t$. 
In view of this uncertainty we investigate a range $U$ from $\sim 6t$--$20t$.

One crucial aspect of the parameters chosen in Eq.~(\ref{eq:onsite3}) requires discussion: 
we recently found that, although in the isolated atom interactions always lead to orbital disproportionation 
(Hund's second rule), in the solid environment this may or may not be the case according to the value of 
$J/U$.\cite{Okamoto04b}
Let us consider minimizing the interaction energy, Eq.~(\ref{eq:onsite3}), with respect to a 
density matrix corresponding to a mean charge density per orbital $n_{\alpha}$ and 
spin density per orbital $m_{\alpha}$. We find [$n(m)_{tot}=\sum_{\alpha} n(m)_\alpha$] 
\begin{eqnarray}
\langle H_{on-site} \rangle [n_{\alpha},m_{\alpha}] = 
\frac{U \left ( n_{tot}^{2}-m_{tot}^{2}\right) }{4}+
\frac{U-5J}{2}\sum_{\alpha>\beta}n_{\alpha}n_{\beta}  \label{Eint} 
+\frac{U-J}{2}\sum_{\alpha>\beta}m_{\alpha}m_{\beta} .
\label{eq:Eee}
\end{eqnarray}
At fixed total density $n_{tot}$ and for a given number of orbitals $n_{orb}$, 
the term $\sum_{\alpha>\beta}n_{\alpha}n_{\beta}$ is maximized by the uniform density 
$n_{\alpha}=n_{tot}/n_{orb}$. 
Thus in a paramagnetic state ($m_{\alpha}=0$ for all $\alpha$) for $J<U/5$ an orbitally 
disproportionated state minimizes the interaction energy, whereas for $J>U/5$ 
a state of uniformly occupied orbitals minimizes the interaction energy. 
For spin polarized states the situation becomes more complicated, 
because the $m_{\alpha}$ and $n_{\alpha}$ are not independent ($m_{\alpha}\leq n_{\alpha}$). 
For fully spin polarized states ($m_{\alpha}=n_{\alpha}$), the condition for 
disproportionation becomes $J<U/3$. 
Therefore, our choice of parameter $J=U/9$ indicates that 
the orbitally disproportionated state becomes stable at large $U$, 
but a larger $J$ would lead to a state with equal orbital occupancy. 

To study the properties of $H_{tot}$ for the three-band model, we employ the
HF approximation replacing terms such as $n_{i \alpha \sigma} n_{i \beta \sigma}$ 
by $n_{i \alpha \sigma} \langle n_{i \beta \sigma} \rangle + 
\langle n_{i \alpha \sigma} \rangle n_{i \beta \sigma}$; 
orbitally off-diagonal expectation values $\langle d_{i \alpha \sigma}^\dag d_{i \beta \sigma} \rangle$ 
of the type considered by Mizokawa\cite{Mizokawa95} and Mochizuki\cite{Mochizuki02} 
are stable only in the presence of a GdFeO$_3$ type distortion which we do not consider. 

\subsection{Single-band model}

In some of our calculations, we consider a simpler model, which is the single-band model. 
The hamiltonian is a simplified representation of the systems studied 
in Ref.~\citenum{Ohtomo02} with the orbital degeneracy neglected. 
We consider nearest-neighbor hopping as in the three-band model. 
Thus, $H_{hop}$ and $H_{on-site}$ are 
\begin{eqnarray}
H_{hop} = - t \sum_{\langle i,j \rangle \sigma} ( d_{i \sigma}^\dag d_{j \sigma} + H.c. ) \label{eq:Hhop1} 
\end{eqnarray}
and
\begin{eqnarray}
H_{on-site}=U \sum_i n_{i \uparrow} n_{i \downarrow}. \label{eq:Honsite1} 
\end{eqnarray} 

The single-band model can be studied by beyond-HF techniques. 
Here we use the dynamical-mean-field method in which 
the basic object of study is the electron Green's function. 
In general, this is given by 
\begin{equation}
G_\sigma(\vec r, \vec {r'} ;\omega )=[\omega +\mu
-H_{hop}-H_{pot}-\Sigma_\sigma (\vec r, \vec {r'} ;\omega )]^{-1}, 
\label{eq:Greenlatt}
\end{equation}%
with the chemical potential $\mu$ and the electron self-energy $\Sigma_\sigma$. 
$H_{pot}$ represents the electrostatic potential from charge +1 counterions placed at La sites. 
In [001] heterostructures with either in-plane translational invariance or 
$N_s$-sublattice antiferromagnetism,  
the Green's function and the self-energy are functions of the variables 
$(z, \eta, z', \eta', \vec{k}_\parallel)$ where $\eta$ and $\eta' (=1,\ldots,N_s)$ label the sublattice 
in layers $z$ and $z'$, respectively, and $\vec{k}_\parallel$ is a momentum in the (reduced) 
Brillouin zone. 
We approximate the self-energy as the sum of a static Hartree term 
$\Sigma_\sigma^H (\vec r, \vec r)$  
arising from the long-ranged part of the Coulomb interaction  
and a dynamical part $\Sigma_\sigma^D (\vec r, \vec {r'}; \omega)$ arising from local fluctuations 
and which we assume that the self-energy is only dependent on layer $z$ and sublattice 
$\eta$.\cite{Schwieger03,Chattopadhyay01a,Chattopadhyay01b} 
Thus, the dynamical part of the self-energy is written as 
\begin{equation}
\Sigma_\sigma^D\Rightarrow \Sigma_\sigma^D (z, \eta; \omega ).  \label{sigd}
\end{equation}
The $z \eta$-dependent self-energy is determined from the solution of 
a quantum impurity model~\cite{Georges96} 
with the mean-field function fixed by the self-consistency condition 
\begin{equation}
G_\sigma^{imp}(z, \eta; \omega)= N_s \int \!\! \frac{d^{2}k_{\parallel}}{(2\pi )^{2}} \,\,
G_\sigma (z, \eta, z, \eta, \vec{k}_{\parallel };\omega ).  \label{sce}
\end{equation}

\subsection{Computational Complexity and Impurity Solvers for DMFT}

In general for the heterostructure with $L$ layers with $N_s$ sublattices, 
one must solve $L \times Ns$ independent impurity models. 
Due to the self-consistency condition [cf. Eq.~(\ref{sce})] 
and to compute the charge density $n_\sigma (z, \eta)=- \int \frac{d\omega}{\pi} 
f(\omega) \mathrm{Im} G_\sigma^{imp}(z, \eta; \omega )$ with $f$ the Fermi distribution function, 
it is required to invert the $(L \times Ns)^2$ Green's function matrix at each momenta and frequency. 
This time consuming numerics restricts the size of the unit cell. 
In our work so far we have considered the commensurate magnetic states with up to two sublattices, 
$N_s =1$ and 2, on each layer and with the charge density independent of the sublattices, i.e., 
paramagnetic (PM), ferromagnetic (FM) states, 
antiferromagnetic (AF) state 
where antiferromagnetic planes with moment alternating from plane to plane, and 
layer-AF state where FM planes with moment alternating from plane to plane. 
Note that the AF state extrapolates to the bulk AF state with magnetic vector 
$\vec q =(\pi,\pi,\pi)$ at $n \rightarrow \infty$. 
By symmetry, the number of quantum impurity models one must solve is reduced to $L$. 
However, solution of the impurity models is a time consuming task, 
and an inexpensive solver is required. 

In this study, we have employed two impurity solvers: 
(i) two-site DMFT\cite{Potthoff01} and (ii) semiclassical approximation (SCA).\cite{Okamoto05PRB} 
The two-site method is a simplified version of the exact-diagonalization method 
approximating a quantum impurity model as a cluster comprised of two orbitals. 
One orbital represents the impurity site which has the same interaction as in the lattice model and 
the other non-correlated (bath) site represents the temporal charge fluctuation.\cite{Potthoff01} 
This method reproduces remarkably accurately the scaling of the quasiparticle weight 
and lower Hubbard band near the Mott transition. 
We have used the two-site method to investigate the dynamical properties of heterostructures 
in the paramagnetic state at $T=0$. 
On the other hand, a small number of bath orbitals, here two, is known to be insufficient 
to describe the thermodynamics correctly~\cite{Okamoto05PRB}. 
To investigate the magnetic behavior at non-zero temperature, 
we apply semiclassical approximation which we have recently developed.\cite{Okamoto05PRB} 
In this approximation, two continuous Hubbard-Stratonovich transformations 
are introduced, coupling to spin- and charge-fields. 
When evaluating the partition function, 
only spin-fields at zero-Matsubara frequency are kept and 
saddle-point approximation is applied for charge-fields at given values of spin-fields. 
This method is found to be reasonably accurate to compute magnetic transition temperature 
because, in most of the correlated electron systems, 
very slow spin-fluctuation becomes dominant near the magnetic transition. 
In contrast to the two-site DMFT, the SCA can not reproduce quasiparticle peak at $\omega=0$. 
This is due to the neglect of quantum fluctuation of Hubbard-Stratonovich fields in the SCA. 
However, the SCA reproduces the spectral function in paramagnetic phase at not-too-low temperature 
and in the strong coupling regime, and in the magnetically ordered phase. 

\section{Hartree-Fock Study of Three-band Model}
\label{sec:HF}

In this section, we present the theoretical results of the realistic three-band-model 
based on the HF approximation. 
Our main focus is on the appearance of various spin and orbital orderings 
different from the bulk ordering. 

\subsection{Phase Diagram} 

\begin{figure}
\begin{center}
\includegraphics[height=6cm,clip]{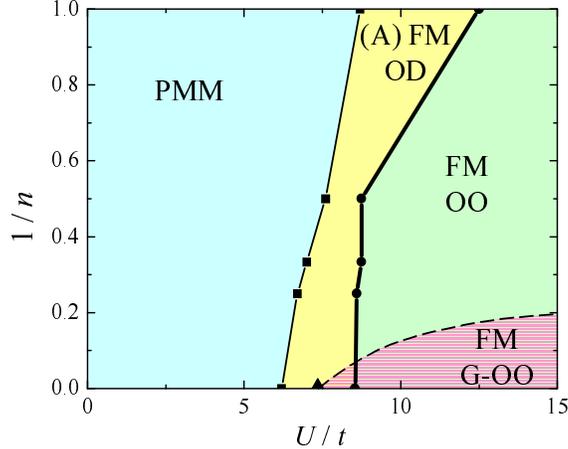}
\end{center}
\caption
{ \label{fig:diagram}
Ground state phase diagram as a function of the on-site Coulomb interaction $U$ 
and inverse of the La layer number $n$ computed by Hartree-Fock approximation. 
PMM: paramagnetic metallic state, 
(A)FM-OD: orbitally disordered state, ferromagnetic at $n=1$ and layer-antiferromagnetic at $n>1$, 
FM-OO: ferromagnetic state with two-dimensional orbital order, 
FM-G-OO: ferromagnetic state with ($\protect\pi,\protect\pi,\protect\pi$) orbital order. 
We take $U^{\prime}=7U/9$, $J=U/9$, and $E_c =0.8$. 
The triangle is the critical $U$ above which ($\protect\pi,\protect\pi,\protect\pi$)
orbital ordering occurs for $n=\infty$. The broken line shows the expected
phase transition to the ($\protect\pi,\protect\pi,\protect\pi$)
antiferromagnetic orbital ordering at finite $n$.}
\end{figure}

Figure~\ref{fig:diagram} shows the calculated spin and orbital phase diagram 
as a function of interaction strength and layer number. 
For reasons of computational convenience in scanning a wide range of parameters, we
considered mainly phases with translation invariance in the $xy$ plane,
however for $n= \infty$ and $n=1$ we also considered an $xy$-plane 
two-sublattice symmetry breaking. We found, in agreement with previous
calculations,\cite{Mizokawa95,Ishihara02,Mochizuki02,Khaliullin02} that the 
fully staggered phase is favored at $n=\infty$, but $xy$-plane symmetry
broken states could not be stabilized in the one layer case. 
We have not yet studied more general symmetry breakings for intermediate layer numbers, 
but the physical arguments presented below strongly suggests that these phases only
occur for larger numbers of layers ($n \agt 6$). 

Let us start with $n=\infty$ limit corresponding to the bulk. 
The comparison to the bulk ordering is subtle. 
In bulk, LaTiO$_3$ exhibits a ($\pi,\pi,\pi$) type antiferromagnetic ordering. 
Theoretical calculations (apparently confirmed by very recent NMR experiments, and x-ray and neutron
diffraction experiments)\cite{Kiyama03,Cwik03} suggest a four-sublattice structure 
which is very close to ($0,0,\pi$)-type orbital ordering\cite{Mochizuki03} 
but differing slightly from the large $U$ ground state studied here.
Stabilizing the observed state apparently requires a lattice distortion not 
included in the model studied here. As $U$ is increased from zero the $n\rightarrow \infty$ 
limit of the model considered here has a phase transition which 
we believe to be of second order to an incommensurate 
antiferromagnetic state with a wave vector which is an extremal spanning
vector of the Fermi surfaces of the bands arising from two of the orbitals
(say $xz,yz$) and which turns out to be very close to ($0,0,\pi$). 
[In fact, for reasons of numerical simplicity we studied ($0,0,\pi$) ordering and
found a very weakly first order transition.] 
Within the model this transition is followed by 
a strongly first order transition to one of a degenerate manifold of states
characterized by ferromagnetic spin order and ($\pi,\pi,\pi$) orbital order
(triangle at $n=\infty$, $U/t \approx 7.5$ in Fig.~\ref{fig:diagram}).
We believe the ($\pi,\pi,\pi$)-orbital spin-ferro state we have found is a 
reasonable surrogate for the actual Mochizuki-Imada state found in experiment. 

Now we turn to the finite $n$ region. 
We observe four phases: a small $U$ phase with no broken symmetry 
[paramagnetic metallic phase (PMM)], and intermediate $U$ phase 
with in-plane translation-invariance spin order, but no orbital order (OD), 
and a large $U$ phase with both spin and orbital order 
[ferromagnetic orbitally ordered phase (FM-OO)]. 
The lower $U$ transition line varies smoothly with layer number
and at $n \rightarrow \infty$ limit it asymptotes to the ($0,0,\pi$) spin ordering in the bulk 
as discussed above. 
The $n=1$ intermediate $U$ phase is ferromagnetic (FM) whereas for $n>1$
the intermediate $U$ phase is antiferromagnetic (AFM). 
The essential reason for those orderings with in-plane translation-invariance 
is that for small $n$ the charge density is spread in the $z$ direction, so no layer has a density near 1.
The larger $U$ transition is essentially independent of layer number for $n>1$. 
Only for sufficiently large $n$, 
is this state preempted by the bulk FM state with ($\pi,\pi,\pi$) OO found at large $U$. 
The essential point is that, for $n<6$, the solution in the large $U$ limit 
consists of several partially filled subbands, which have effectively 
minimized their interaction energy but which gain considerable kinetic 
energy from motion in the $xy$ plane. Breaking of $xy$ translation symmetry 
would reduce this kinetic energy gain without much decreasing the already 
saturated interaction energy while $z$-direction kinetic energy is quenched 
by the confining potential. 
Therefore, although computational difficulties have prevented us from precisely 
locating the FM-OO to bulk phase boundary we expect the dashed line in 
Fig.~\ref{fig:diagram} is a reasonable estimate. 
For completeness, we have also shown the $n \rightarrow \infty$ limit of the FM-OD to 
FM-OO phase boundary, calculated by suppressing the bulk phase. 

\subsection{Density Distribution and Metallic Edge} 
We now turn to the spatial distribution of the charge density its relation to the metallic behavior. 
These are less sensitive to details of the ordered pattern and to the precise values of parameters. 

\begin{figure}[b]
\begin{minipage}[t]{8.25cm}
\begin{center}
\begin{tabular}{c}
\includegraphics[height=6cm,clip]{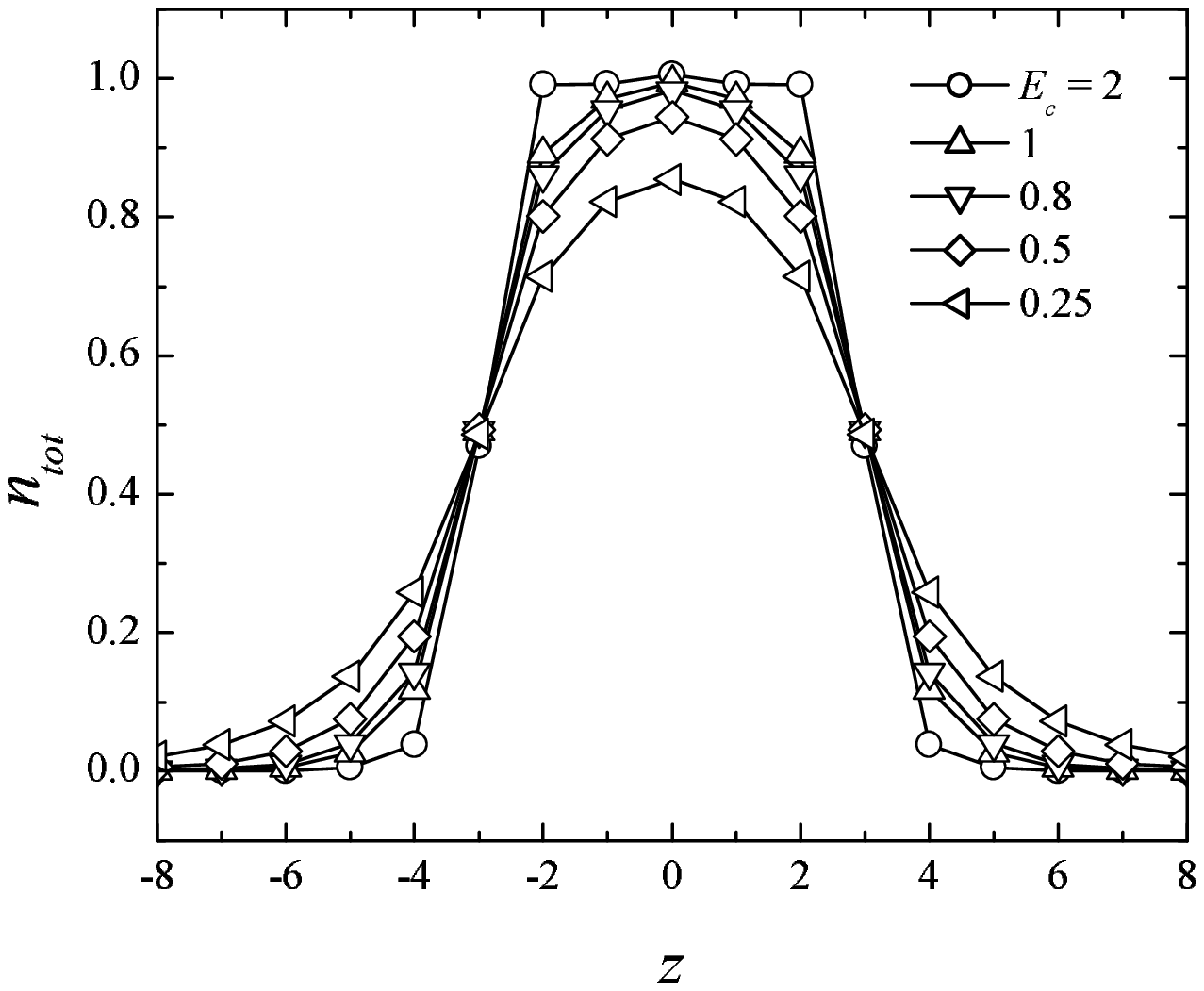}
\end{tabular}
\caption
{ \label{fig:Ecdep}
Total density profile of $n=6$ heterostructure with different values of $E_c$ indicated. 
$U/t=10$, and a paramagnetic state is assumed. 
$E_c=2$ and 0.25 correspond to the dielectric constant $\varepsilon \sim 6$ 
and $\sim 48$, respectively, with $t \sim 0.3$~eV and $a \sim 3.9$~\AA. 
}
\end{center}
\end{minipage} \quad
\begin{minipage}[t]{8.25cm}
\begin{center}
\begin{tabular}{c}
\includegraphics[height=6cm,clip]{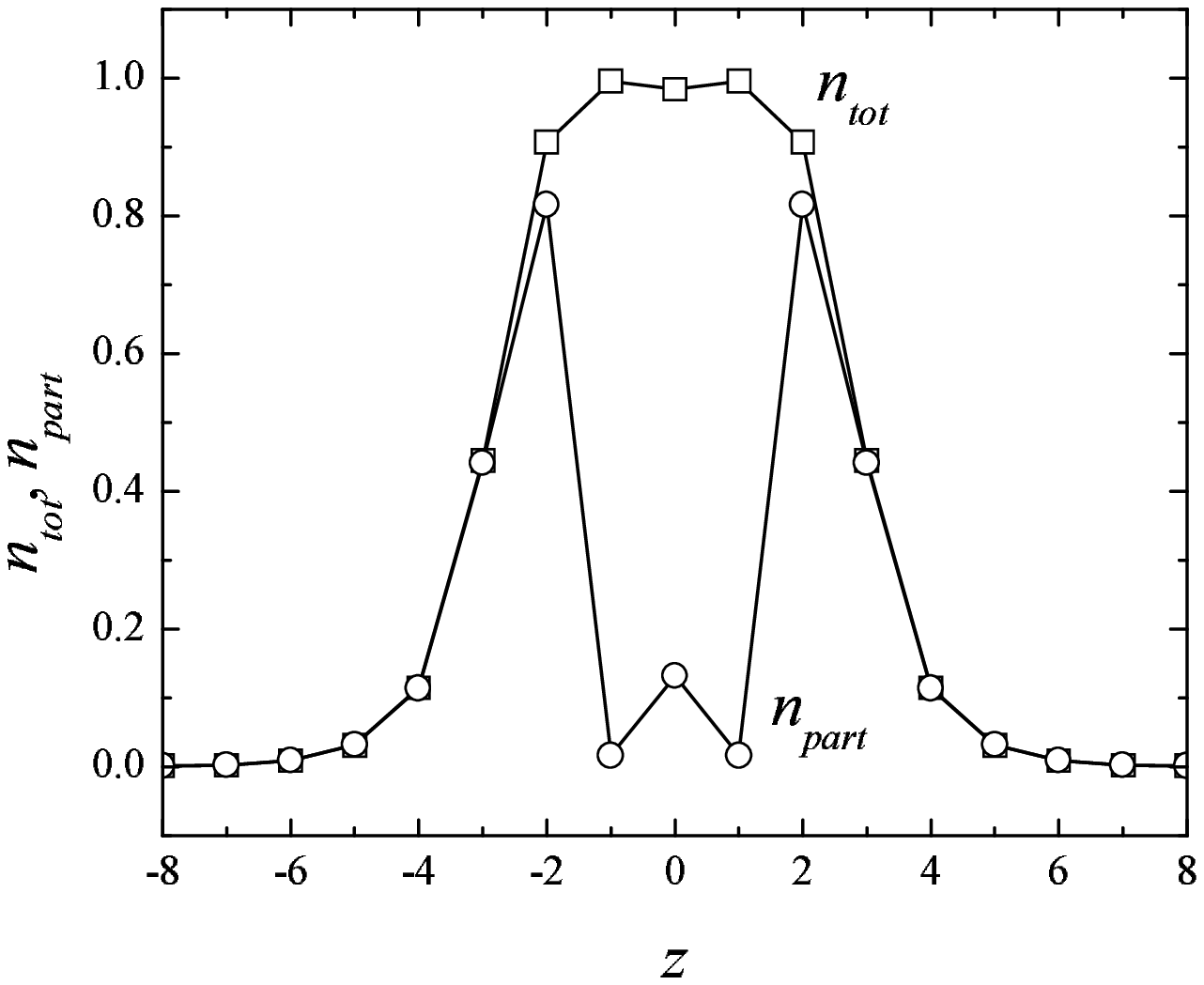}
\end{tabular}
\caption
{ \label{fig:NtotNpart}
Total density profile (open squares) and 
density in partially occupied bands $n_{part}$ (open circles) of heterostructure with 
$n=6$, $U/t=10$ and $E_c = 0.8$ 
calculated in orbitally ordered ferromagnetic phase. 
Charge +1 counterions are placed at $z=\pm 0.5, \pm 1.5, \pm 2.5$, 
so the electronic ($B$) sites are at the integer values of $z$. 
}
\end{center}
\end{minipage}
\end{figure} 

Typical results for density distribution are presented in Figs.~\ref{fig:Ecdep} and \ref{fig:NtotNpart} 
for the $n=6$ heterostructure with $U/t=10$. 
Charge +1 counterions are placed at $z=\pm 0.5, \pm 1.5, \pm 2.5$, thus, 
the regions at $|z|< (>)3$ correspond to LaTiO$_3$ (SrTiO$_3$). 
Figure~\ref{fig:Ecdep} shows charge density distributions in the paramagnetic phase 
for different values of $E_c=e^2/(\varepsilon at)$. 
Parameter values $E_c=2$ and 0.25 correspond to the dielectric constant $\varepsilon \sim 6$ 
and $\sim 48$, respectively, with $t \sim 0.3$~eV and $a \sim 3.9$~\AA. \
It is seen that the density distribution does not depend sensitively on $E_c$, 
thus, neither does the HF result for the phase diagram. 
However, it is expected that $E_c$ affects the boundary between phases with and without in-plane symmetry 
because the stability of the in-plane symmetry broken phase is sensitive to 
how the density is close to 1 which is required for the bulk like ordering. 
This issue is discussed in detail for the single-band model in Ref.~\citenum{Okamoto05}. 
Figure~\ref{fig:NtotNpart} presents density profile (open squares) of 
the orbitally ordered ferromagnetic state with $E_c=0.8$. 
It is seen that difference from the paramagnetic state shown in Fig.~\ref{fig:Ecdep} is small. 
At the interface between the two regions $|z|=3$, 
we observe about three-unit-cell-wide crossover regime where the density drops from $\sim 1$ to $\sim 0$. 
This crossover region turned out to support metallic behavior, thus we term it a {\em metallic edge}. 

Within the HF method we use here, 
physics of the metallic edge is manifested as follows. 
There are many bound-state solutions (wave functions decaying as 
$|z|$ is increased away form the heterostructure), 
whose dispersion in the in-plane direction gives rise to subbands. 
For thin heterostructures, all subbands are partially filled 
(implying metallic behavior in the heterostructure plane) 
whereas for thick heterostructures (in ordered phases), 
some subbands are fully filled and some are partly filled. 
The fully filled subbands have $z$-direction wave functions implying charge 
density concentrated in the middle of the heterostructure, whereas the 
partially filled bands have charge concentrated near the edges. 
The occupancy of partially filled bands is computed and presented as open circles in 
Fig.~\ref{fig:NtotNpart}. 
We observe partially filled bands corresponding to metallic 
behavior at the crossover regime. 

Summarizing this section, 
we applied the HF approximation to model heterostructures comprised of three-band Hubbard model. 
The ground-state phase diagram shows that thin heterostructures exhibit spin and orbital orderings 
different from the bulk ordering. 
The charge density at the edge region where the density drops from $\sim 1$ to $\sim 0$ 
is dominated by the partially filled bands indicating metallic behavior in this region, 
although the two component compounds are insulating in bulk. 
These results demonstrate the concept of {\em electronic reconstruction} in correlated-electron 
heterostructures. 

As will be discussed in the next section, 
details of phase boundaries and ordering patterns will be changed when beyond-HF methods are applied. 
Furthermore, values of on-site interaction in real materials may be changed 
by changing the thickness and changing the screening effect. 
However, the general conclusion that non-bulk phases occur and that an interesting series of phase transitions
may take place as the layer thickness is varied are robust. 

\section{Dynamical-mean-field Study of Single-band Model}
\label{sec:DMFT}

This section presents DMFT studies of single-band heterostructures. 
First, we present results for the spatial variation of dynamical properties of electrons 
such as single-particle spectral functions. 
It is shown that the quasiparticle appearing at interfaces between 
Mott-insulator region and band-insulator region is only moderately correlated and 
gives rise to metallic behavior. 
Second, we discuss magnetic orderings in the heterostructures 
which are different from the bulk ordering. 
Theoretical results confirm the important results obtained by the HF approximation shown in 
the previous section, i.e., 
different phases in thin heterostructures than in the bulk, and metallic edge, 
but show the significant improvement providing the reasonable estimates of the transition temperature 
and new insights into the spatial variation of the order parameter. 

\subsection{Heavy Quasiparticle and Metallic Edge}
\label{subsec:2siteDMFT}

Here, we apply the two-site DMFT to the single-band heterostructures 
to investigate the dynamical properties of heterostructures. 
We consider paramagnetic states at $T=0$. 
This is because we are interested in the crossover behavior between a Mott-insulating state, 
where the insulating behavior purely comes from the electron correlation, 
and the HF method can not be applied, to the others. 

\begin{figure}
\begin{minipage}{16cm}
\begin{center}
\begin{tabular}{c}
\includegraphics[height=8.5cm,clip]{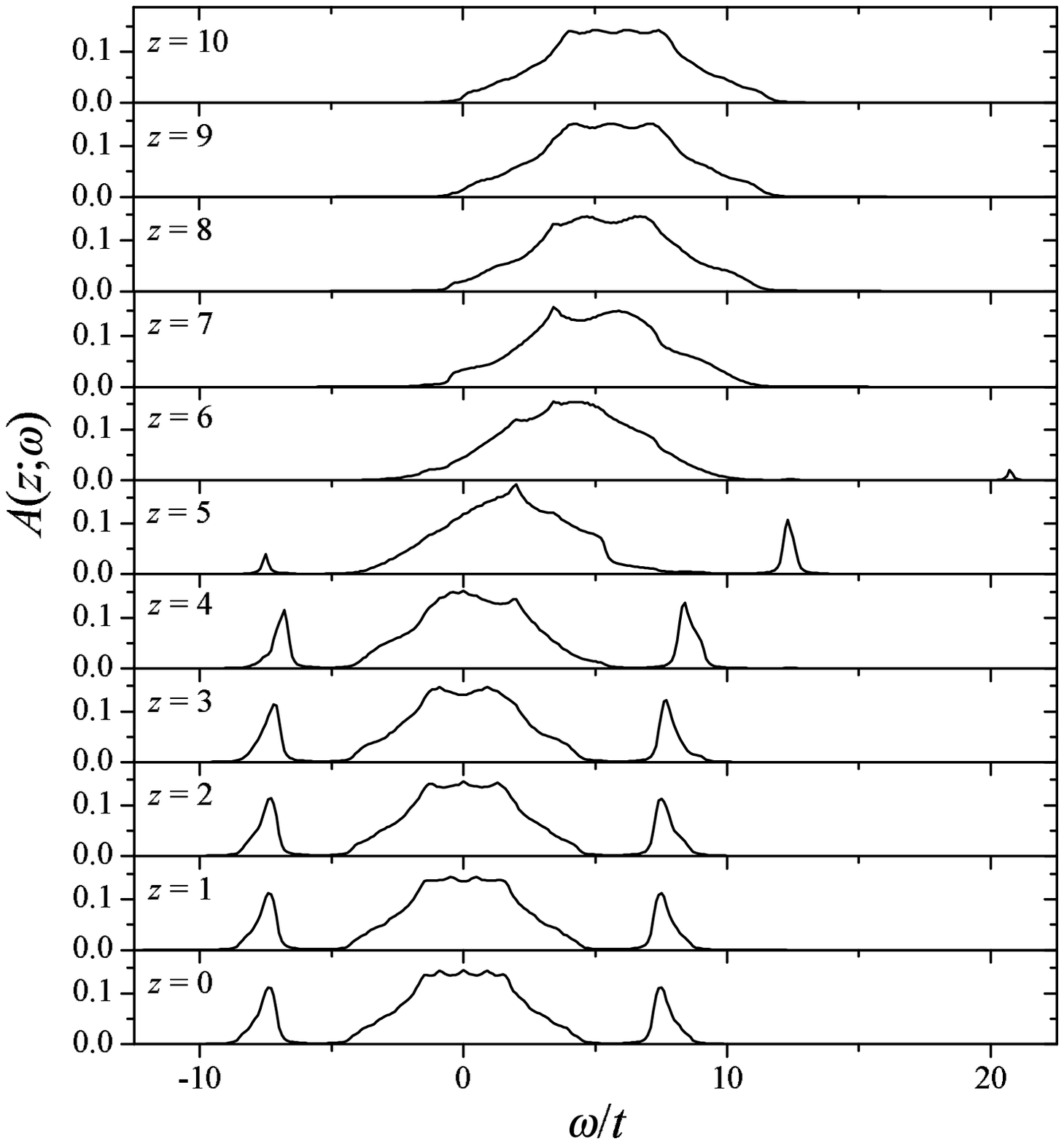}
\qquad
\includegraphics[height=8.5cm,clip]{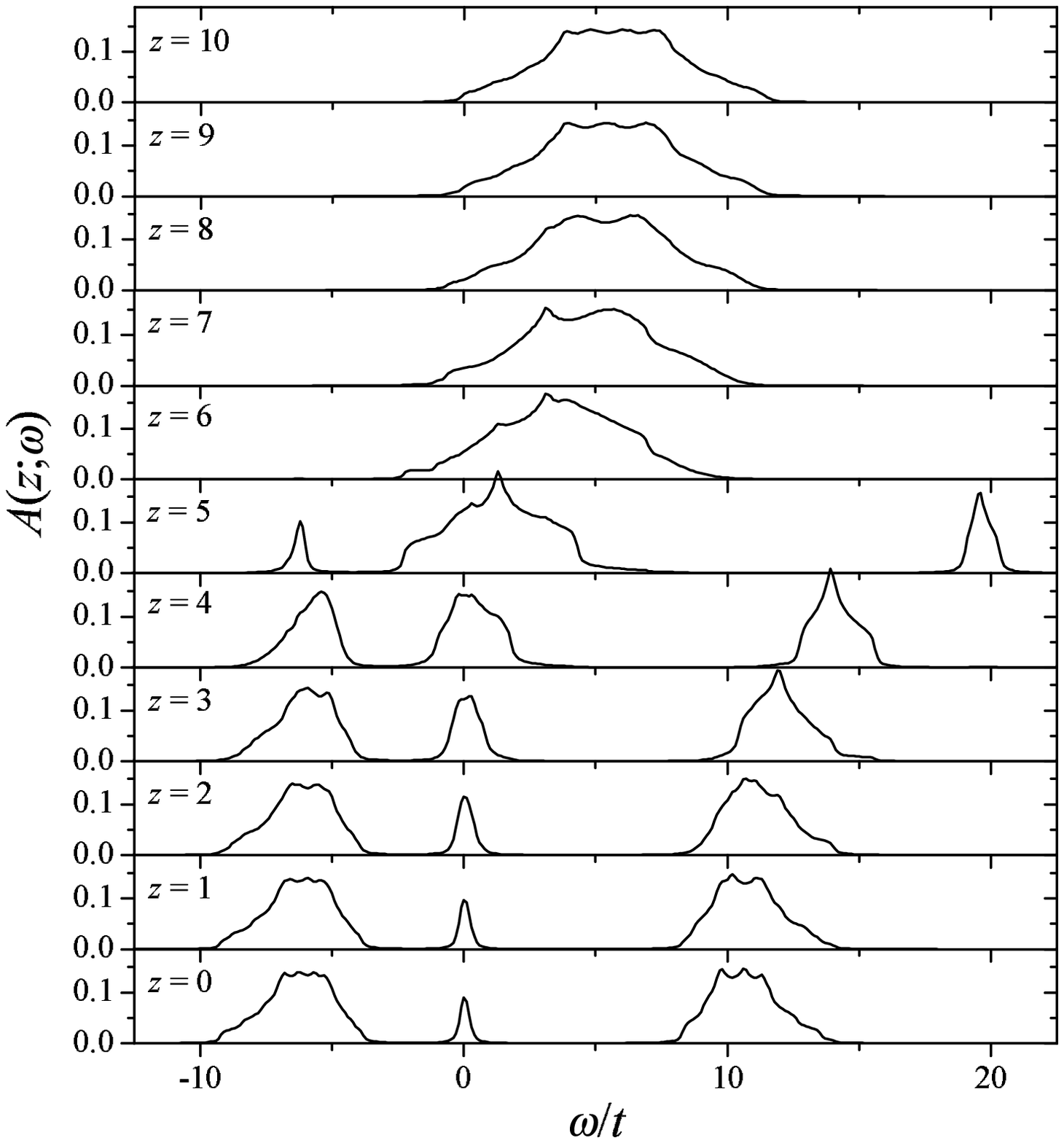}
\end{tabular}
\end{center}
\end{minipage}
\caption
{ \label{fig:spectra}
Layer-resolved spectral function calculated for $10$-layer
heterostructure with $E_c =0.8$. 
Left panel: $U/t=10$, right panel: $U/t=16$. 
Charge +1 counterions are placed at $z=\pm 0.5, \ldots ,\pm 4.5$ so the
electronic ($B$) sites are at integer values of $z$.}
\end{figure}

One of the most useful observable to see the dynamical property of the correlated electron 
is the single particle spectral function. 
The spectral functions are in principle measurable in photoemission
or scanning tunneling microscopy. 
Numerical results for the layer-resolved spectral function 
$A(z;\omega)=-\frac{1}{\pi }\int \frac{d^{2}k_{\parallel }}{(2\pi )^{2}}%
\mathrm{Im}G(z,z,\vec{k}_{\parallel };\omega +i0^{+})$ for a $10$-layer
heterostructure with different values of $U$ are presented in Fig.~\ref{fig:spectra}. 
The dimensionless parameter for the long-ranged Coulomb interaction is $E_c=0.8$. 
The left panel shows results for the weak coupling ($U/t=10$), 
and the right panel for the strong coupling ($U/t=16$ 
about 10\% greater than the critical value which drives the Mott transition in a bulk system described by 
$H$ with $n = \infty$). 
The critical value for the bulk Mott transition is estimated to be $U_c/t \approx 14.7$ 
by the two-site DMFT. 
Outside the heterostructure ($|z| \gg 6$), 
the spectral function is essentially identical in form to that of the free
tight-binding model $H_{band}$ for both the weak coupling and strong coupling results. 
The electron density is negligible, as can be seen from the fact that almost all of 
the spectral function lies above the chemical potential. 
As one approaches the heterostructure ($|z|=6$), the spectral function begins to broaden. 
For the weak coupling case, spectral functions at $|z|<6$ are also quite similar to the results of 
the HF analysis except for tiny peaks outside of central quasiparticle band 
corresponding to the upper- and lower-Hubbard bands. 
On the other hand for the strong coupling case, 
spectral weight around $\omega =0$ begins to decrease rapidly and the characteristic strong 
correlation structure of lower and upper Hubbard bands with a central quasiparticle peak begins to form. 
The sharp separation between these features is an artifact of the two-site DMFT 
[as is, we suspect, the shift in energy of the upper (empty state) Hubbard band for $z=4,5$]. 
Experience with bulk calculations suggests that the existence of three features and the weight in the
quasiparticle region are reliable. 
Towards the center of the heterostructure, the weight in the quasiparticle band becomes very small,
indicating nearly insulating behavior. 
For very thick heterostructures, we find the weight approaches $0$ exponentially.

\begin{figure}
\begin{center}
\includegraphics[height=6cm,clip]{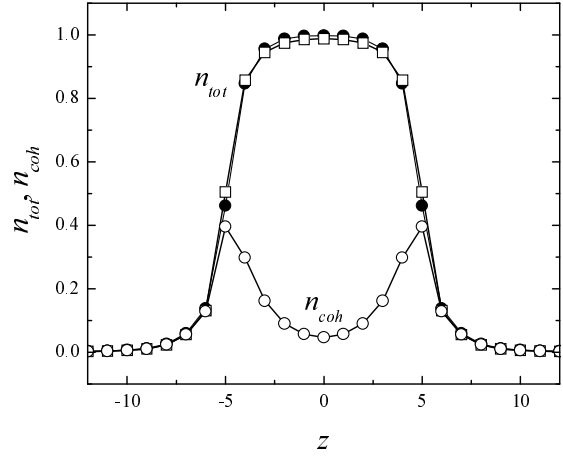}
\end{center}
\caption
{ \label{fig:density}
Total charge density (open squares) and charge density from the 
coherent part near the Fermi level (open circles). 
For comparison, the total charge density calculated by applying the HF approximation to 
the Hamiltonian is shown as filled symbols. 
The parameters are the same as in the right panel of Fig.~\ref{fig:spectra}.}
\end{figure}

The behavior shown in Fig.~\ref{fig:spectra} is driven by the variation in
density caused by leakage of electrons out of the heterostructure region.
Figure~\ref{fig:density} shows as open squares the numerical results for the charge-density
distribution $n_{tot}(z)$ for the heterostructure whose photoemission spectra are
shown in Fig.~\ref{fig:spectra}. One sees that in the center of the
heterostructure ($z=0$) the charge density is approximately $1$ per site,
and that there exists an edge region, of about three-unit-cell width, over
which the density drops from $\sim 1$ to $\sim 0$. The overall charge
profile is determined mainly by the self-consistent screening of the Coulomb
fields which define the heterostructure, and is only very weakly affected by
the details of the strong on-site correlations (although the fact that the
correlations constrain $n_{tot}<1$ is obviously important). To show this, we have
used the HF approximation to recalculate the charge profile: the
results are shown as filled circles in Fig.~\ref{fig:density} and are seen
to be almost identical to the DMFT results.

In order to study the metallic behavior associated with the quasiparticle subband, 
we computed the charge density from the quasiparticle bands $n_{coh}$ 
by integrating $A(z;\omega )$ from $\omega =0$ down to the first point at which $A(z;\omega)=0$. 
Results are shown as open circles in Fig.~\ref{fig:density}. 
It is obvious that these near-Fermi-surface states 
contain a small but non-negligible fraction of the total density, suggesting 
that edges should display relatively robust metallic behavior. 
The results represent a significant correction to the HF calculation shown in 
Fig.~\ref{fig:NtotNpart}, which leads, in the edge region, to a metallic quasiparticle
density essentially equal to the total density.
This is because the HF approximation does not give mass renormalization. 
Calculations, not shown here, of the dispersion of the quasiparticle subbands find that 
the mass renormalization is, to a good approximation, 
$m^*/m \sim n_{tot}/n_{coh}$. 

\subsection{Magnetic Ordering at Finite Temperature}
\label{subsec:SCA}

Now, we turn to the magnetic ordering of the single-band heterostructures at non-zero temperature. 
The HF approximation provides a very poor approximation to the behavior in this region. 
We also discuss differences between DMFT and HF. 

\begin{figure}[b]
\begin{center}
\includegraphics[height=6cm,clip]{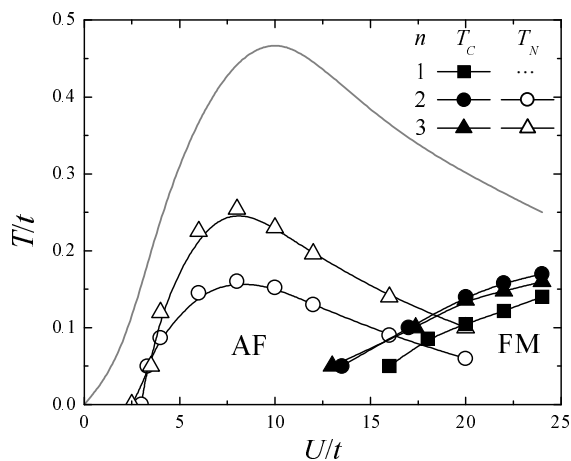}
\end{center}
\caption
{ \label{fig:PD_DMFT}
Magnetic transition temperatures of heterostructures with various thicknesses $n$ 
as functions of interaction strength. $E_c=0.8$. 
Filled symbols: Ferromagnetic Curie temperature $T_C$, open symbols: 
Antiferromagnetic N{\' e}el temperature $T_N$. 
Note that the $n=1$ heterostructure does not exhibit a N{\'e}el phase. 
Note that, where both phases are locally stable, 
the phase with higher $T_C (T_N)$ has the lower free energy. 
For comparison, $T_N$ for bulk AF ordering computed by the same method is shown 
as the light solid line. 
}
\end{figure} 

Figure~\ref{fig:PD_DMFT} shows our calculated phase diagram in 
the interaction-temperature plane for heterostructures with various thicknesses. 
The one-layer heterostructure is PM at weak to moderate interactions, and FM at strong interactions. 
The two- and three-layer heterostructures are AF at weak to intermediate interaction, 
and become FM at stronger interactions with almost the same $T_C$ for $n=2$ and 3. 
The phase diagram displays regions where both $T_C$ and $T_N >0$ (FM and AF both locally stable); 
in these regions the phase with the higher transition temperature has the lower free energy and 
is the one which actually occurs. 
HF studies of the single-orbital model find a layer-AF phase. 
This phase is not found in our DMFT analysis. 

Note that an antiferromagnetic ordering in a two-dimensional system occurs only at $T=0$. 
The non-zero $T_N$ is an artifact of the mean-field nature of the DMFT. 
In a real system, slow antiferromagnetic fluctuations are expected to develop for temperatures below the 
$T_N$ determined by the DMFT.  
As will be shown below, the ferromagnetism is an interface effect. 
We expect that at large $U$, very thick heterostructures will be AF in the center, 
but with a FM surface layer. 
Antiferromagnetic N{\'e}el temperature $T_N$ is found to be strongly dependent on the layer thickness; 
it increases with the increase of layer thickness.  
Note that $T_N$'s are substantially reduced from the bulk value, 
$T_N \sim 6t^2/U$ at strong-coupling regime (see a light solid line in Fig.~\ref{fig:PD_DMFT}), 
due to the smaller charge density per site. 
Further, the in-plane symmetry broken phase is found to become unstable 
by increasing the dielectric constant and broadening the charge density profile. 
(For details, see Ref.~\citenum{Okamoto05}.) 

\begin{figure}
\begin{minipage}[t]{8.25cm}
\begin{center}
\begin{tabular}{c}
\includegraphics[height=8.5cm,clip]{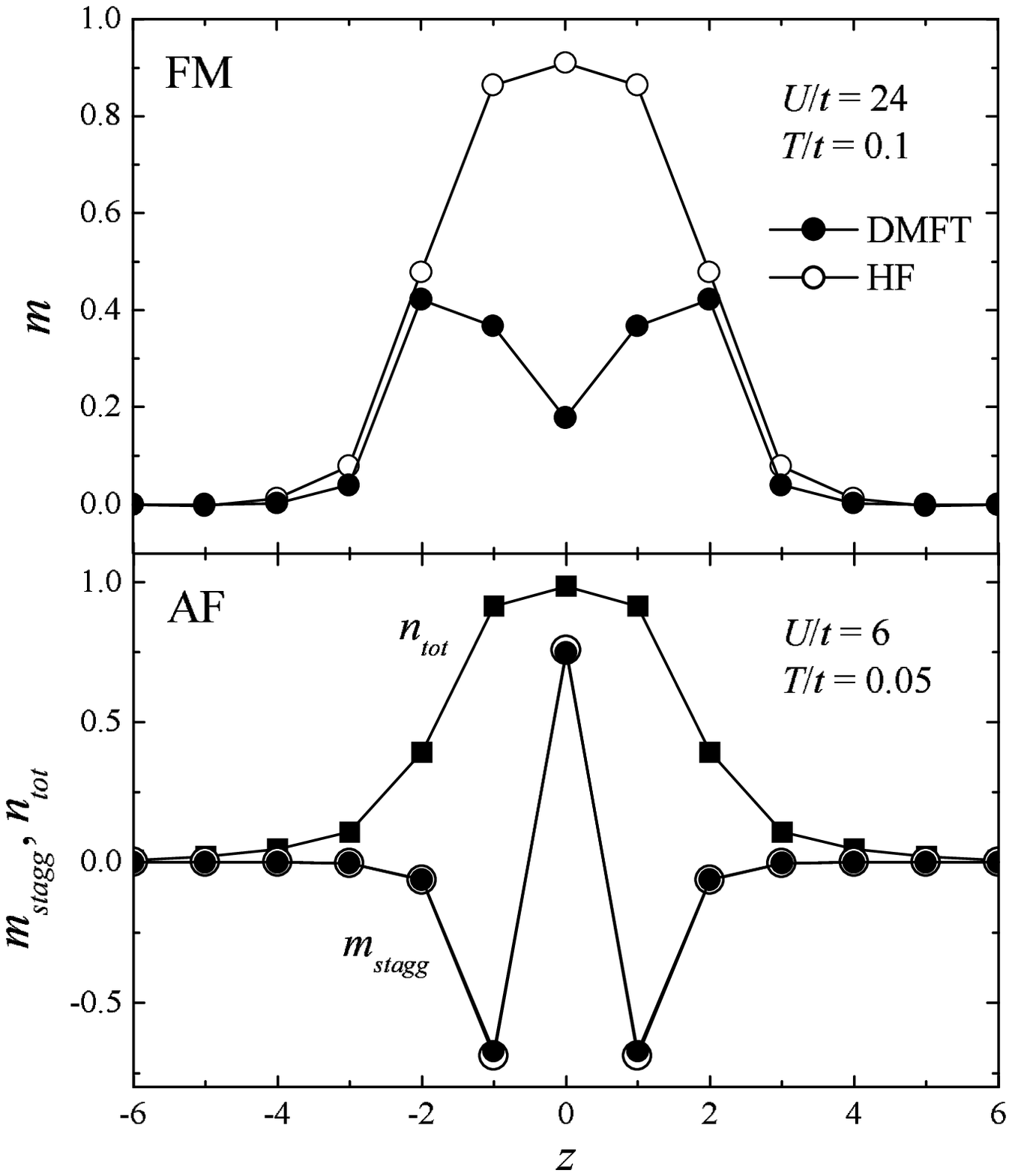}
\end{tabular}
\caption
{ \label{fig:NMvsZ}
Magnetization density of 4-layer heterostructure. 
Counterions are placed at $z= \pm 0.5, \pm 1.5$. $E_c=0.8$. 
Upper panel: Magnetization $m$ in a FM state for $U/t=24$ and $T/t=0.1$. 
Lower panel: In-plane staggered magnetization $m_{stagg}$ in an AF state for 
$U/t=6$ and $T/t=0.05$. 
Filled (open) circles are the results by the DMFT (HF). 
For comparison, charge density $n_{tot}$ computed by the DMFT is also shown in 
the lower panel (filled squares). 
Note that the staggered magnetization in the outer layers ($|z| \ge 2$), 
and the outermost layers ($|z|=1$) have the same sign.}
\end{center}
\end{minipage} \quad
\begin{minipage}[t]{8.25cm}
\begin{center}
\begin{tabular}{c}
\includegraphics[height=8.5cm,clip]{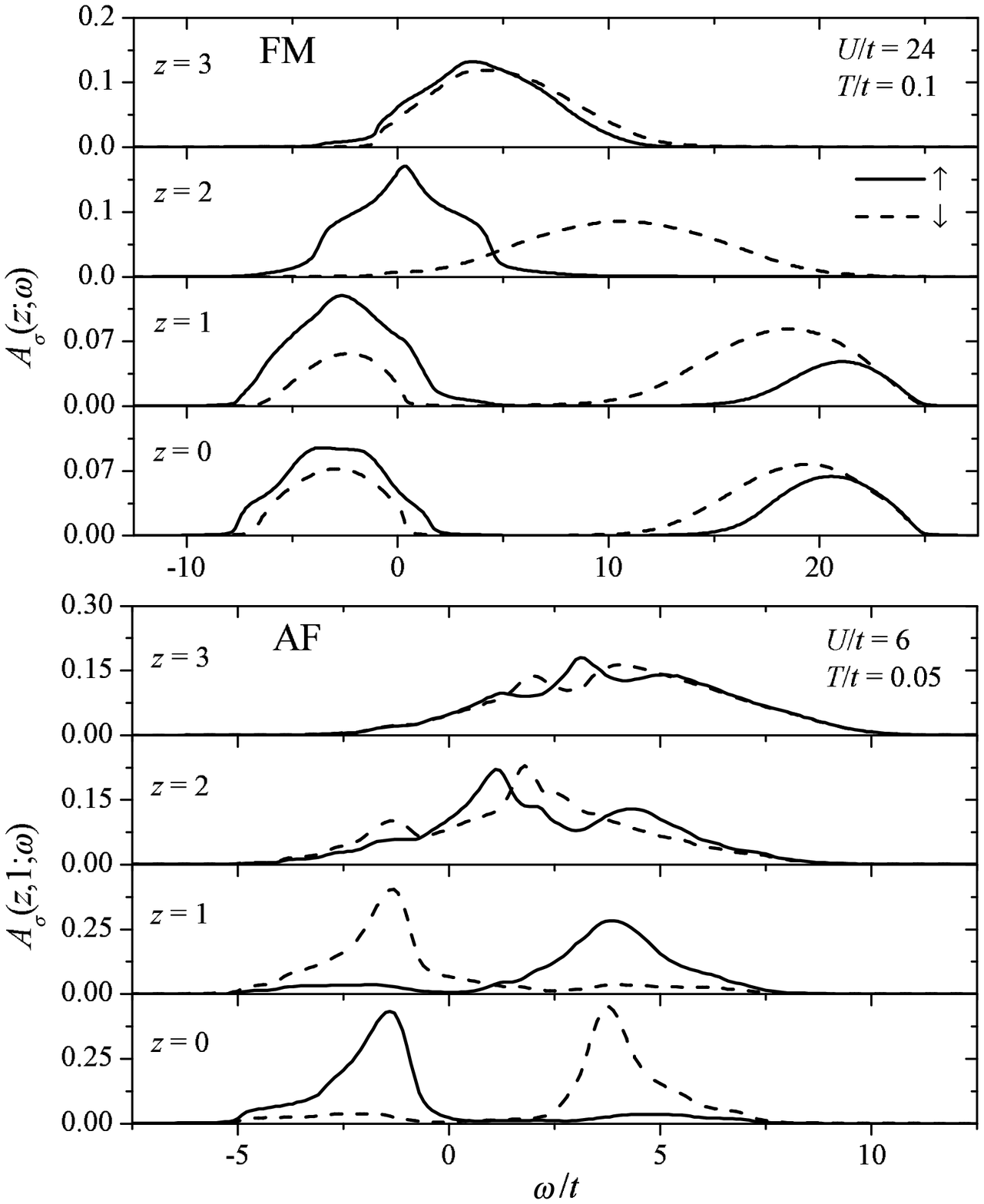}
\end{tabular}
\caption
{ \label{fig:spect_SCA}
Layer- and sublattice-resolved spectral functions as functions of real frequency $\omega$
for 4-layer heterostructure. $E_c=0.8$. 
Upper panel: Ferromagnetic state at $U/t=24$ and $T/t=0.1$. 
Lower panel: Antiferromagnetic state at $U/t=6$ and $T/t=0.05$. Sublattice $\eta =1$. 
Solid (broken) lines are for up (down) spin electrons. 
For sublattice $\eta =2$ in AF state, up and down electrons are interchanged.}
\end{center}
\end{minipage}
\end{figure}

We now turn to the spatial variation of the magnetization density. 
As examples, numerical results for a 4-layer heterostructure 
(counterions at $z=\pm 0.5$ and $\pm1.5$) with $E_c=0.8$ are presented in Fig.~\ref{fig:NMvsZ}. 
The upper panel of Fig.~\ref{fig:NMvsZ} shows the magnetization in the FM state. 
In the DMFT (filled circles), only the layers near the interfaces ($|z| \sim 2$) have large polarization 
and inner layers in the heterostructure have small moments. 
This explains the weak $n$-dependence of $T_C$ of thick heterostructures 
(see the upper panel of Fig.~\ref{fig:PD_DMFT}). 
In the HF (open circles), all layers in the heterostructure are highly polarized. 
In contrast, in AF states, the in-plane staggered magnetization computed by the DMFT and the HF 
agree well as shown in the lower panel of Fig.~\ref{fig:NMvsZ}. 
For comparison, the total charge density is also plotted (filled squares). 
The in-plane staggered magnetization is large only at inner layers where the charge density is close to 1. 
Note that the staggered magnetization in the outer layers ($|z| \ge 2$) has the same sign 
as in the layers at $|z|=1$ indicating that the outer layers are not intrinsically magnetic. 

The spin distributions presented in Fig.~\ref{fig:NMvsZ} can be understood from the 
single-particle spectral functions. 
In Fig.~\ref{fig:spect_SCA} are presented the DMFT results for 
the layer- and sublattice-resolved spectral functions 
$A_\sigma(z, \eta; \omega)= - \frac{1}{\pi} \mathrm{Im} G_\sigma^{imp}(z,\eta; \omega +i0^{+})$ 
for the FM (upper panel) and the AF (lower panel) states 
of 4-layer heterostructure with the same parameters as in Fig.~\ref{fig:NMvsZ}. 
These quantities can in principle be measured by 
spin-dependent photoemission or scanning tunneling microscopy. 
As noticed in Sec.~\ref{subsec:2siteDMFT}, 
spectral function outside of the heterostructure ($|z| \gg 2$) is essentially identical 
to that of the free tight-binding model $H_{band}$, and electron density is negligibly small. 
Approaching the interfaces ($|z|=2$), 
the spectral function shifts downwards and begins to broaden. 
In the FM case, magnetic ordering is possible only near the interface ($|z| \sim 2$) 
where the charge density is intermediate. 
Inside the heterostructure ($|z| < 2$), clear Hubbard gap exists due to the large $U$ 
and uniform polarization is hard to achieve. 
On the contrary, high charge density is necessary to keep the staggered magnetization 
in the AF case as seen as a difference between up and down spectra in the lower panel of 
Fig.~\ref{fig:spect_SCA}. 

Summarizing this section, 
by applying the DMFT to the single-band heterostructure, 
we confirmed two important results obtained in the previous section on HF analysis: 
thin heterostructures show different magnetic orderings and an approximately three-unit-cell-wide 
interface region becomes metallic. 
Using the DMFT, we obtained new information about correlated-heterostructure behaviors: 
metallic edge involves quasiparticle bands with moderate mass enhancement, 
ferromagnetic Curie temperature is estimated to be low compared with the bulk N{\'e}el temperature, 
ferromagnetic moment is concentrated in the interface region where there is intermediate charge density 
while antiferromagnetic staggered moment is concentrated in the region where the density is close to 
the one for the bulk Mott insulating compound. 
Thus, {\em electronic reconstruction} is a generic property of the correlated-electron heterostructures. 

\section{Conclusion} 
\label{sec:conclusion}

This paper summarizes our theoretical studies of the electronic properties of a 
{\em correlated heterostructure} model involving $n$ layers of 
a material which in bulk is a Mott insulator, embedded in an infinite band insulator. 
The specific features of the model we study were chosen to reproduce 
the LaTiO$_3$/SrTiO$_3$ heteostructure system studied by Ohtomo \textit{et al.}, 
but we hope that our results will shed light also on the 
more general question of the physics of interfaces between 
strongly correlated and weakly correlated systems.

A crucial feature of the experimental LaTiO$_3$/SrTiO$_3$ system studied by
Ohtomo \textit{et al.} is the almost perfect lattice match between the two
systems. These authors argued that this implies that the only difference
between the Mott insulating and band insulating regions arises from the
different charge of the La$^{3+}$ and Sr$^{2+}$; in particular the crystal
structure and atomic positions are expected to remain relatively constant
throughout the heterostructure. 
Of course, the asymmetry present at the LaTiO$_3$/SrTiO$_3$ interface must induce 
some changes in atomic positions: a TiO$_6$ octahedron is negatively charged, 
and so if it sits between a Sr plane and a La plane it will be attracted to the latter, 
and also distorted, because the positively charged Ti will tend to move in the opposite
direction. 
(When one uses $\varepsilon=1$, 
the electrostatic force exerted by a LaO plane on the neighboring O$^{2-}$ ion 
is estimated to be $\approx 20$~eV/\AA \ !) 
The experimentally determined Ti-Ti distances shown in Fig.~1 of
Ref.~\citenum{Ohtomo02}, along with the distortion in that paper,
suggests that the changes in Ti-Ti distance are negligible. 
In this circumstance, changes in O position along the Ti-O-Ti bond change hoppings
only in second order. 
We therefore neglected these effects and assumed that the electronic hoppings and 
interaction parameters remain invariant across the heterostructure. 
However, we emphasize that properly accounting for the 
effect of atomic rearrangements inevitably present at surface and interface is crucial. 
We further note that lattice distortions appear to be important
in stabilizing the observed bulk state, but may be pinned in a heterostructure. 
Extending our results to include these effects is an important open problem. 

In the calculations reported here, the heterostructure is defined only by the
difference (+3 vs +2) of the La and Sr charge. The calculated electronic
charge density is found to be controlled mainly by electrostatic effects 
(ionic potentials screened by the electronic charge distribution). 
Results, such as shown in Figs.~\ref{fig:Ecdep} and \ref{fig:NtotNpart} 
for the three-band model
or Fig.~\ref{fig:density} for the single-band model, are representative of results 
obtained for a wide range of on-site interaction $U$, long-ranged Coulomb interaction $E_c$, 
and different theoretical methods. 
We find generally that significant leakage of charge into the band insulator region occurs. 
The width of the transition region must depend on the relative
strength of the $z$-direction hoppings and the confining potential. 
For the parameters studied here, the transition region where the charge density 
changes from the value for the bulk Mott-insulator $\sim 1$ to the one for 
the band-insulator $\sim 0$ is about three layers. 
The spreading of the electronic charge controls the electronic properties of the heterostructure. 

Using Hartree-Fock and dynamical-mean-field approximations, 
we demonstrated that spin (and orbital) orderings in thin heterostructures differ from that in the bulk. 
This behavior originates from the lower charge density in the heterostructure 
due to the leakage of electrons from the Mott-insulating region to the band-insulating region. 
The dynamical properties of correlated-electron heterostructures were 
calculated using the dynamical-mean-field method. 
Our results show how the electronic behavior evolves from the weakly correlated 
to the strongly correlated regions, and in particular, confirms the existence 
of an approximately three-unit-cell-wide crossover region in which a system, 
insulating in bulk, can sustain metallic behavior. 
We found that even in the presence of very strong bulk correlations, the metallic edge behavior
displays a moderate mass renormalization. 
In Ref.~\citenum{Okamoto04PRB2}, we have also discussed 
how the magnitude of the renormalization is affected by the spatial
structure of the quasiparticle wave function and determined how this
renormalization affects physical observables such as the optical conductivity. 
Further, in Ref.~\citenum{Okamoto05}, we have 
investigated how changes in the charge density distribution affects the magnetic transition temperature. 

Being based on the theoretical studies, 
we would like to propose a new and important concept in the correlated-electron 
interface problem: {\em electronic reconstruction} which means 
{\em electronic phase behavior at interfaces differs from that in the bulk}. 
{\em Electronic reconstruction} is originally proposed by Hesper {\it et al}. 
to describe the rearrangement of the electronic charge in the presence of a polar surface of 
C$_{60}$ films.\cite{Hesper00} 
We suggest this phrase be applied more generally to the electronic phase behavior. 
This contrasts with the ordinary {\em lattice reconstruction} implying the appearance of 
the different lattice structure at the interface than in the bulk 
of un-correlated or weakly correlated materials. 

Finally, 
important future directions for research include re-examination of the
phase diagram of three-band model using beyond Hartree-Fock techniques, 
and generalization of the results presented here to more realistic cases. 
As a first step towards more realistic systems, 
we have started the analysis of a theoretical model on manganites superlattices 
using the DMFT method.\cite{Lin05} 
We also note that experiments measuring the single-particle excitations 
in titanates based superlattices by means of photoemission spectroscopy 
has already started.\cite{Takizawa05} 

\acknowledgments     
 
We acknowledge fruitful discussions with H. Hwang, A. Ohtomo, H. Monien, M. Potthoff. 
G. Kotliar, P. Sun, J. Chakhalian, B. Keimer, R. Ramesh, and A. Fujimori. 
This research was supported by JSPS (S.O.) and the DOE under Grant No. ER 46169 (A.J.M.).



\begin{thebibliography}{1}
\bibitem{Imada98}M.~Imada, A.~Fujimori, and Y.~Tokura, ``Metal-insulator transitions,''
{\em Rev. Mod. Phys.} \textbf{70}, pp.~1039-1263, 1998.

\bibitem{Tokura00}Y.~Tokura and N.~Nagaosa, ``Orbital physics in transition-metal oxides,'' 
{\em Science} \textbf{288}, pp.~462-468, 2000.


\bibitem{Moore04}R.~G.~Moore, J.~Zhang, S.~V.~Kalinin, Ismail, A.~P.~Baddorf, R.~Jin, 
D.~G.~Mandrus, and E.~W.~Plummer, 
``Surface dynamics of the layered ruthenate Ca$_{1.9}$Sr$_{0.1}$RuO$_4$,''
{\em Phys. Status Solidi (b)} {\bf 241}, pp.~2363-2366, 2004.

\bibitem{Potthoff99a}M.~Potthoff and W.~Nolting, 
``Surface metal-insulator transition in the Hubbard model,''
{\em Phys. Rev. B} \textbf{60}, pp.~2549-2555, 1999. 

\bibitem{Potthoff99b}M.~Potthoff and W.~Nolting, 
``Metallic surface of a Mott insulator-Mott insulating surface of a metal,''
{\em Phys. Rev. B} \textbf{60}, pp.~7834-7849, 1999. 

\bibitem{Schwieger03}S.~Schwieger, M.~Potthoff, and W.~Nolting, 
``Correlation and surface effects in vanadium oxides,'' 
{\em Phys. Rev. B} {\bf 67}, 165408, 2003. 

\bibitem{Liebsch03}A.~Liebsch, 
``Surface versus bulk Coulomb correlations in photoemission spectra of SrVO$_3$ and CaVO$_3$,'' 
{\em Phys. Rev. Lett.} {\bf 90}, 096401, 2003. 

\bibitem{Hesper00} R.~Hesper, L.~H.~Tjeng, A.~Heeres, and G.~A.~Sawatzky, 
``Photoemission evidence of electronic stabilization of polar surfaces in K$_3$C$_{60}$,''
{\em Phys. Rev. B} \textbf{62}, pp.~16046-16055, 2000.

\bibitem{Maiti01}K.~Maiti, D.~D.~Sarma, M.~J.~Rozenberg, I.~H.~Inoue, H.~Makino, O.~Goto, M.~Pedio, and R.~Cimino, 
``Electronic structure of Ca$_{1-x}$Sr$_x$VO$_3$: A tale of two energy scales,'' 
{\em Europhys. Lett.} {\bf 55}, pp.~246-252, 2001. 

\bibitem{Okamoto04b}S.~Okamoto and A.~J.~Millis, 
``Electron-lattice coupling, orbital stability, and the phase diagram of Ca$_{2-x}$Sr$_x$RuO$_4$,''
{\em Phys. Rev. B} {\bf 70}, 195120, 2004. 

\bibitem{Ahn99}C.~H.~Ahn, S.~Gariglio, P.~Paruch, T.~Tybell, L.~Antognazza, and J.-M.~Triscone, 
``Electrostatic modulation of superconductivity in ultrathin GdBa$_2$Cu$_3$O$_{7-x}$ films,''
{\em Science} \textbf{284}, pp.~1152-1155, 1999.

\bibitem{Gariglio02}S.~Gariglio, C.~H.~Ahn, D.~Matthey, and J.-M.~Triscone, 
``Electrostatic tuning of the hole density in NdBa$_2$Cu$_3$O$_{7- \delta}$ 
films and its effect on the Hall response,''
{\em Phys. Rev. Lett.} \textbf{88}, 067002, 2002. 

\bibitem{Schneider02}C.~W.~Schneider, S.~Hembacher, G.~Hammerl, R.~Held, A.~Schmehl, A.~Weber, T.~Kopp, 
and J.~Mannhart, 
``Electron Transport through YBa$_2$Cu$_3$O$_{7-\delta}$: Grain Boundary Interfaces between 4.2 and 300 K,''
{\em Phys. Rev. Lett.} \textbf{92}, 257003, 2002.

\bibitem{Ohtomo02} A.~Ohtomo, D.~A.~Muller, J.~L.~Grazul, and H.~Y.~Hwang,
``Artificial charge-modulationin atomic-scale perovskite titanate superlattices,'' 
{\em Nature} \textbf{419}, pp.~378-380, 2002.

\bibitem{Izumi99}M.~Izumi, Y.~Murakami, Y.~Konishi, T.~Manako, M.~Kawasaki, and Y.~Tokura, 
``Structure characterization and magnetic properties of oxide superlattices 
La$_{0.6}$Sr$_{0.4}$MnO$_3$/La$_{0.6}$Sr$_{0.4}$FeO$_3$,'' 
{\em Phys. Rev. B} \textbf{60}, pp.~1211-1215, 1999.

\bibitem{Izumi01}M.~Izumi, Y.~Ogimoto, Y.~Konishi, T.~Manako, M.~Kawasaki, and Y.~Tokura, 
``Perovskite superlattices as tailored materials of correlated electrons,'' 
{\em Mat. Sci. Eng. B} \textbf{84}, pp.~53-57, 2001. 

\bibitem{Biswas00}A.~Biswas, M.~Rajeswari, R.~C.~Srivastava, Y.~H.~Li, T.~Venkatesan, R.~L.~Greene, and A.~J.~Millis, 
``Two-phase behavior in strained thin films of hole-doped manganites,'' 
{\em Phys. Rev. B} \textbf{61}, pp.~9665-9668, 2000.  

\bibitem{Biswas01}A.~Biswas, M.~Rajeswari, R.~C.~Srivastava, T.~Venkatesan, R.~L.~Greene, Q.~Lu, 
A.~L.~de Lozanne, and A.~J.~Millis,
``Strain-driven charge-ordered state in La$_{0.67}$Ca$_{0.33}$MnO$_3$,'' 
{\em Phys. Rev. B} \textbf{63}, 184424, 2001.

\bibitem{Warusawithana03}M.~P.~Warusawithana, E.~V.~Colla, J.~N.~Eckstein, and M.~B.~Weissman, 
``Artificial dielectric superlattices with broken inversion symmetry,'' 
{\em Phys. Rev. Lett.} {\bf 90}, 036802, 2003. 

\bibitem{Bowen03}M.~Bowen, M.~Bibes, A.~Barthelemy, J.~P.~Contour, A.~Anane, Y.~Lemaitre, and A.~Fert, 
``Nearly total spin polarization in La$_{2/3}$Sr$_{1/3}$MnO$_3$ from tunneling experiments,''
{\em Appl. Phys. Lett.} \textbf{82}, pp.~233-235, 2003. 

\bibitem{Nakagawa05}N.~Nakagawa, M.~Asai, Y.~Mukunoki, T.~Susaki, and H.~Y.~Hwang, 
``Magnetocapacitance and exponential magnetoresistance in manganite-titanate heterojunctions,''
{\em Appl. Phys. Lett.} \textbf{86}, 082504, 2005. 

\bibitem{Mannhart05}J.~Mannhart, 
``Nano-magnetism at interfaces in high-temperature superconductors?'' 
talk at the workshop on {\em Nanoscale Fluctuations in Magnetic and Superconducting Systems}, 
Dresden, 2005. 

\bibitem{Potthoff95}M.~Potthoff and W.~Nolting, 
``Surface magnetism studied within the mean-field approximation of the Hubbard model,''
{\em Phys. Rev. B} {\bf 52}, pp.~15341-15354, 1995.

\bibitem{Matzdorf00}R.~Matzdorf, Z.~Fang, Ismail, J.~Zhang, T.~Kimura, Y.~Tokura, K.~Terakura, and E.~W.~Plummer, 
``Ferromagnetism stabilized by lattice distortion at the surface of the $p$-wave superconductor Sr$_2$RuO$_4$,''
{\em Science} \textbf{289}, pp.~746-748, 2000. 

\bibitem{Calderon99}M.~J.~Calder\'on, L.~Brey, and F.~Guinea, 
``Surface electronic structure and magnetic properties of doped manganites,'' 
{\em Phys. Rev. B} {\bf 60}, pp.~6698-6704, 1999. 

\bibitem{Fang00}Z.~Fang, I.~V.~Solovyev, and K.~Terakura, 
``Phase diagram of tetragonal manganites,'' 
{\em Phys. Rev. Lett.} \textbf{84}, pp.~3169-3172, 2000.

\bibitem{Tinte03}S.~Tinte, K.~M.~Rabe, and D.~Vanderbilt, 
``Anomalous enhancement of tetragonality in PbTiO$_3$ induced by negative pressure,''
{\em Phys. Rev. B} {\bf 68}, 144105, 2003. 

\bibitem{Bungaro04}C.~Bungaro and K.~M.~Rabe, 
``Epitaxially strained [001]-(PbTiO$_3$)$_1$(PbZrO$_3$)$_1$ superlattice and PbTiO$_3$ from first principles,''
{\em Phys. Rev. B} 69, 184101, 2004. 

\bibitem{Ederer05}C.~Ederer and N.~A.~Spaldin, 
``Influence of strain and oxygen vacancies on the magnetoelectric properties of multiferroic bismuth ferrite,''
{\em Phys. Rev. B} {\em 71}, 224103, 2005. 

\bibitem{Popovic05}Z.~S.~Popovic and S.~Satpathy, 
``Wedge-shaped potential and Airy-function electron localization in oxide superlattices,''
{\em Phys. Rev. Lett.} {\bf 94}, 176805, 2005. 

\bibitem{Freericks04}J.~K.~Freericks, 
``Dynamical mean-field theory for strongly correlated inhomogeneous multilayered nanostructures,''
{\em Phys. Rev. B} \textbf{70}, 195342, 2004. 

\bibitem{Saitoh95} T.~Saitoh, A.~E.~Bocquet, T.~Mizokawa, and A.~Fujimori,
``Systematic variation of the electronic structure of $3d$ transition-metal compounds,''
{\em Phys. Rev. B} \textbf{52}, pp.~7934-7938, 1995.

\bibitem{Okamoto04Nat}S.~Okamoto and A.~J.~Millis, 
``Electronic reconstruction at an interface between a Mott insulator and a band insulator,''
{\em Nature} {\bf 428}, pp.~630-633, 2004.

\bibitem{Okamoto04PRB1}S.~Okamoto and A.~J.~Millis, 
``Theory of Mott insulator-band insulator heterostructures,'' 
{\em Phys. Rev. B} {\bf 70}, 075101, 2004. 

\bibitem{Georges96}A.~Georges, G.~Kotliar, W.~Krauth, and M.~J.~Rozenberg, 
``Dynamical mean-field theory of strongly correlated fermion systems and the limit of infinite dimensions,'' 
{\em Rev. Mod. Phys.} \textbf{68}, pp.~13-125, 1996.

\bibitem{Okamoto04PRB2}S.~Okamoto and A.~J.~Millis,
``Spatial inhomogeneity and strong correlation physics: 
a dynamical mean field study of a model Mott-insulator--band-insulator heterostructure,''
{\em Phys. Rev. B} {\bf 70}, 241104(R), 2004. 

\bibitem{Okamoto05}S.~Okamoto and A.~J.~Millis,
``Magnetic ordering in a model Mott-insulator--band-insulator heterostructure,''
cond-mat/0506172. 

\bibitem{Maclean79} D.~A.~Maclean, H.-N.~Ng, and J.~E.~Greedan, 
``Crystal structures and crystal chemistry of the $RE$TiO$_3$ perovskites: $RE$ = La, Nd, Sm, Gd, Y,''
{\em J. Solid State Chem.} \textbf{30}, pp.~35-44, 1979.

\bibitem{Sunstrom92}J.~E.~Sunstrom IV, S.~M.~Kauzlarich, and  P.~Klavins
``Synthesis, structure, and properties of lanthanum strontium titanate 
(La$_{1-x}$Sr$_x$TiO$_3$) ($0 \le x \le 1$),''
{\em Chem. Mater.} \textbf{4}, pp.~346-353, 1992.

\bibitem{Sakudo71}T.~Sakudo and H.~Unoki, 
``Dielectric properties of SrTiO$_3$ at low temperatures,'' 
{\em Phys. Rev. Lett.} \textbf{26}, pp.~851-853, 1971; 
K.~A.~M{\"u}ller and H.~Burkard, 
``SrTiO$_3$: An intrinsic quantum paraelectric below 4 K,''
{\em Phys. Rev. B} \textbf{19}, pp.~3593-3602, 1973.

\bibitem{Fujitani95} H.~Fujitani and S.~Asano, 
``Full-potential band calculations on YTiO$_3$ with a distorted perovskite structure,''
{\em Phys. Rev. B} \textbf{51}, pp.~2098-2102, 1995. 

\bibitem{Slater54}J.~C.~Slater and G.~F.~Koster, 
``Simplified LCAO method for the periodic potential problem,'' 
{\em Phys. Rev.} \textbf{94}, pp.~1498-1524, 1954.

\bibitem{Mizokawa95} T.~Mizokawa and A.~Fujimori, 
``Unrestricted Hartree-Fock study of transition-metal oxides: 
Spin and orbital ordering in perovskite-type lattice,''
{\em Phys. Rev. B} \textbf{51}, pp.~12880-12883, 1995.

\bibitem{Okimoto95} Y.~Okimoto, T.~Katsufuji, Y.~Okada, T.~Arima, and Y.~Tokura, 
``Optical spectra in (La,Y)TiO$_3$: 
Variation of Mott-Hubbard gap features with change of electron correlation and band filling,''
{\em Phys. Rev. B} \textbf{51}, pp.~9581-9588, 1995.

\bibitem{Mochizuki02} M.~Mochizuki, 
``Spin and orbital states and their phase transitions of the perovskite-type Ti oxides: Weak coupling approach,''
{\em J. Phys. Soc. Jpn.} \textbf{71}, pp.~2039-2047, 2002.

\bibitem{Chattopadhyay01a}A.~Chattopadhyay and A.~J.~Millis, 
``Theory of transition temperature of magnetic double perovskites,'' 
{\em Phys. Rev. B} {\bf 64}, 024424, 2001. 

\bibitem{Chattopadhyay01b}A.~Chattopadhyay, S.~Das Sarma, and A.~J.~Millis, 
``Transition temperature of ferromagnetic semiconductors: A dynamical mean field study,''
{\em Phys. Rev. Lett.} {\bf87}, 227202, 2001.

\bibitem{Potthoff01}M.~Potthoff, 
``Two-site dynamical mean-field theory,'' 
{\em Phys. Rev. B} {\bf 64}, 165114, 2001. 

\bibitem{Okamoto05PRB}S.~Okamoto, A.~Fuhrmann, A.~Comanac, and A.~J.~Millis,  
``Benchmarkings for a semiclassical impurity solver for dynamical-mean-field theory: 
Self-energies and magnetic transitions of the single-orbital Hubbard model,'' 
{\em Phys. Rev. B} {\bf 71}, 235113, 2005. 

\bibitem{Ishihara02} S.~Ishihara, T.~Hatakeyama, and S.~Maekawa, 
``Magnetic ordering, orbital ordering, and resonant x-ray scattering in perovskite titanates,'' 
{\em Phys. Rev. B} \textbf{65}, 064442, 2002.

\bibitem{Khaliullin02} G.~Khaliullin and S.~Okamoto, 
``Quantum behavior of orbitals in ferromagnetic titanates: Novel orderings and excitations,''
{\em Phys. Rev. Lett.} \textbf{89}, 167201, 2002.

\bibitem{Kiyama03} T.~Kiyama and M.~Itoh, 
``Presence of 3$d$ quadrupole moment in LaTiO$_3$ studied by $^{47,49}$Ti NMR,''
{\em Phys. Rev. Lett.} \textbf{91}, 167202, 2003.

\bibitem{Cwik03} M.~Cwik, T.~Lorenz, J.~Baier, R.~M{\"u}ller, G.~Andre,
F.~Bouree, F.~Lichtenberg, A.~Freimuth, R.~Schmitz, E.~M{\"u}ller-Hartmann, and M.~Braden, 
``Crystal and magnetic structure of LaTiO$_3$: Evidence for nondegenerate $t_{2g}$ orbitals,''
{\em Phys. Rev. B} \textbf{68}, 060401(R), 2003.

\bibitem{Mochizuki03} M.~Mochizuki and M.~Imada, 
``Orbital-spin structure and lattice coupling in $R$TiO$_3$ where $R$ = La, Pr, Nd, and Sm,''
{\em Phys. Rev. Lett.} \textbf{91}, 167203, 2003.

\bibitem{Lin05}C.~Lin, A.~J.~Millis, and S.~Okamoto, 
``Dynamical mean field study of Manganite superlattices,'' 
(in preparation). 

\bibitem{Takizawa05}M.~Takizawa, T.~Wadati, M.~Kobayashi, S.~Tanaka, S.~Yagi, 
S.~Hashimoto, T.~Yoshida, A.~Fujimori, S.~Chikamatsu, H.~Kumigashira, 
S.~Ojima, K.~Shibuya, S.~Mihara, T.~Ohonishi, M.~Lippmaa, M.~Kawasaki, H.~Koinuma  
(private communication). 


\end{thebibliography}
\end{document}